\documentclass[aps,twocolumn,showpacs,superscriptaddress]{revtex4}

\usepackage[dvips]{graphicx}
\usepackage{amsmath}
\usepackage{amsfonts}
\usepackage{amssymb} 

\renewcommand{\textrm}{\rm}

\newcommand{\ovl}{\overline}
\newcommand{\td}{\tilde}
\newcommand{\C}{^*}
\newcommand{\T}{^\mathrm{T}}
\newcommand{\TC}{^\mathrm{T*}}
\newcommand{\Tr}{\mathop{\mathrm{Tr}}}

\newcommand{\vecb}[1]{{\bf{#1}}}
\newcommand{\upd}{{\mathrm{d}}}
\newcommand{\dhat}{\hat{d}}
\newcommand{\lhat}{\hat{l}}
\newcommand{\khat}{\hat{k}}
\newcommand{\mhat}{\hat{m}}
\newcommand{\nhat}{\hat{n}}

\newcommand{\Ohat}{\hat{O}}
\newcommand{\shat}{\hat{s}}
\newcommand{\lvechat}{\hat{\vecb l}}
\newcommand{\mvechat}{\hat{\vecb m}}
\newcommand{\nvechat}{\hat{\vecb n}}
\newcommand{\rvechat}{\hat{\vecb r}}
\newcommand{\dvechat}{\hat{\vecb d}}

\newcommand{\wvechat}{\hat{\vecb w}}

\newcommand{\vvec}{{\vecb v}}
\newcommand{\Svec}{{\vecb S}}
\newcommand{\svechat}{\hat{\vecb s}}
\newcommand{\kvechat}{{\hat{\vecb k}}}

\newcommand{\Ovechat}{{\hat{\vecb O}}}
\newcommand{\xvec}{{\vecb x}}

\newcommand{\xvechat}{\hat{\vecb x}}
\newcommand{\yvechat}{\hat{\vecb y}}
\newcommand{\zvechat}{\hat{\vecb z}}

\newcommand{\Hvec}{\vecb H}

\newcommand{\avec}{\vecb a}

\newcommand{\jvec}{\vecb j}

\newcommand{\Deltavec}{\boldsymbol{\Delta}}
\newcommand{\sigmavec}{\boldsymbol{\sigma}}
\newcommand{\epsilonvec}{\boldsymbol{\epsilon}}

\newcommand{\omegavechat}{\hat{\boldsymbol{\omega}}}

\newcommand{\orbv}{\Ovechat}
\newcommand{\orbc}{\Ohat}
\newcommand{\iu}{{\mathrm{i}}}
\newcommand{\im}{\mathop{\textrm{Im}}}
\newcommand{\re}{\mathop{\textrm{Re}}}
\newcommand{\rotaxis}{\omegavechat}

\newcommand{\spin}{\textrm{spin}} 
\newcommand{\rmit}{} 
\newcommand{\refrmit}{\emph} 

\begin{document}

\title{Equilibrium Simulations of 2D Weak Links in 
$p$-wave Superfluids}


\author{Janne K. Viljas}
\affiliation{Low Temperature Laboratory, Helsinki University of
Technology, \\P.O.Box 2200, FIN-02015 HUT, Finland}
\author{Erkki V. Thuneberg}
\affiliation{Low Temperature Laboratory, Helsinki University of
Technology, \\P.O.Box 2200, FIN-02015 HUT, Finland}
\affiliation{Department of Physical Sciences, P.O.Box 3000,\\
FIN-90014 University of Oulu, Finland}



\begin{abstract}
A two-dimensional Ginzburg-Landau theory of weak links in a 
$p$-wave superfluid is presented. First we consider the symmetry
properties of the energy functionals, and their relation to the
conserved supercurrents which play an essential role in the weak link 
problem. In numerical studies, we use the A and B phases
of superfluid $^3$He. The phases on the two sides of the weak link can
be chosen separately, and very general soft degrees of freedom
may be imposed as boundary conditions. We study all four inequivalent 
combinations of A and B which are possible for a hole in a planar
wall, including weak links with a pinned A-B interface. In all cases, some illustrative
current-phase relations (CPR's) are calculated and the critical
currents are mapped. Phase diagrams covering the relevant phase space
in zero magnetic field are constructed. The numerical methods are also
described in some detail. 
%
\end{abstract}


\maketitle

\section{INTRODUCTION} \label{s.intro}

Recent experimental observations\cite{berkeley,paris} of the
Josephson effect in weak links of superfluid $^3$He left theorists
with some interesting problems. One of these was related to the interpretation 
of the ``$\pi$ states'', local minima of the Josephson energy at
phase differences other than $0$ (mod $2\pi$). 
Inspired by these findings, some theoretical work was 
done.\cite{viljas,yippi,viljas2,nishida,zhang,hatakenaka} By now there exists a
reasonably good understanding of how such $\pi$ states could be 
obtained in small ``pinhole'' contacts,\cite{yippi,viljas2,nishida}
in arrays of apertures via the anisotextural
effect,\cite{viljas,viljas2} or in single large apertures
in the Ginzburg-Landau (GL) regime.\cite{viljas,thuneberg88} However, large
physically relevant parts of the parameter space remain unexplored.
In this paper we study the GL regime in more detail, and attempt to
bridge the gap between the pinhole and the large-aperture limits.

In Ref.\ \onlinecite{viljas} we studied a three-dimensional (3D) circular
aperture between two bulk volumes of B phase $^3$He by solving the GL
equations in a full 3D lattice. With this
calculation we were able to demonstrate that a local minimum of energy
exists at phase differences close to $\pi$, which is associated with a
separate ``$\pi$ branch'' in the current-phase relation. This solution
was never clearly found to be a global minimum of energy, and no jumps to it from the
strongly hysteretic ``$0$ branches'' could be observed. Therefore it appeared to
be experimentally inaccessible. This calculation was very restricted,
however, since the bulk order parameters were fixed to be equal on the
two sides. In the present calculation we are able to impose much more
general boundary conditions on the soft degrees of freedom of the
order parameters, and to use the A phase in addition
to the B phase. On the other hand, we have not attempted to carry out
the calculations in 3D, but rather we use a two-dimensional (2D) model of a long
slit-shaped weak link in a planar wall.\cite{thuneberg88} 
Nevertheless, as our experience obtained with
the previous calculation of Ref.\ \onlinecite{viljas} suggests, we may
expect very similar effects to exist in all weak links, regardless of
their shape. In the A phase this is not quite so, as will be discussed
below, because the critical current can even vanish in some very
symmetrical situations.

Since this article is to appear in the proceedings of a winter school,
our approach is in many ways tutorial. Furthermore, since the involved
work is largely numerical, special attention is paid to explaining the
computational methods. Throughout, we shall only deal with thermodynamic
equilibrium, and the numerics are thus mostly related to a minimization of the GL
free-energy functional on a square lattice. Our relaxation methods are
actually quite standard ones, but have never been
presented in detail in the specific context of $^3$He. 

In Section 2 we discuss the Josephson coupling on a formal 
level, and remind how non-sinusoidal energy-phase relations can be
obtained via a perturbative approach in $^3$He and unconventional
superconductors alike. From Section 3 onward we turn to the 
case of $^3$He, and first review some basic issues and notations.
We relate the symmetries of the free energy
functional to the conserved supercurrents, and
review the GL theory briefly. In particular, we consider spin supercurrents on an equal
footing with the mass supercurrents everywhere. 
Section 4 introduces our 2D weak link model, with the 
divisions into numerical and asymptotic regions, and discusses
physical ways to control the boundary conditions. 
Finally, an analysis of the four different phase combinations is given
in Section 5, and some examples of current-phase relations 
(CPR's) are presented. All critical currents and phase diagrams are 
summarized in Section 6. 
In Section 7 we end with some conclusions and 
discussion. The asymptotic solutions far from the weak link
are developed in Appendix A, and an analysis of the
numerical method is presented in Appendix B.

\section{DC JOSEPHSON EFFECT REVISITED} \label{s.dc}

The superfluid state of a BCS superfluid/superconductor is described by
a gap matrix, or a pairing potential. This $2\times2$ matrix has the
form
\begin{equation}
\Delta_{\alpha\beta}(\kvechat)=[\Delta_s\iu\sigma_y+\Deltavec
\cdot\sigmavec \iu\sigma_y]_{\alpha\beta}.
\end{equation}
Here $\Delta_s$ and $\Deltavec$ are the singlet and triplet pairing
amplitudes, respectively, and their argument $\kvechat$ parametrizes
points on the Fermi surface.
From the Pauli principle
$\Delta_{\alpha\beta}(\kvechat)=-\Delta_{\beta\alpha}(-\kvechat)$
it follows that $\Delta_s$ and $\Deltavec$ have even parity ($s,d,g,\ldots$) and odd
parity ($p,f,h,\ldots$) in $\kvechat$, respectively. 
The gap matrix also serves as the order parameter of the superfluid.

The dc (or \emph{equilibrium}) Josephson effect results from a
coupling of two superfluids (left, $L$, and right, $R$) through a
weak link. Due to this coupling, the free energy of the system is
generally changed by an amount $F_J$ which we call the \emph{Josephson
coupling energy}. 
We assume $F_J$ to be a function of the bulk values $\Delta^{L,R}$
of the order parameters. 
 From the order parameters we
separate out phase factors by writing 
$\Delta^{L,R}=\Delta^{L,R}_0\exp(\iu\phi^{L,R})$. Because of global
gauge invariance $F_J$ should independent of the global phase. The
dependence of
$F_J$ on the phase difference
$\Delta\phi=\phi^R-\phi^L$ is  called the
energy-phase relation (EPR). Since the phase factors are $2\pi$
periodic, so is the EPR: $F_J(\Delta\phi+2\pi)=F_J(\Delta\phi)$. As is
well known, the derivative of EPR with respect to $\Delta\phi$ gives
the current-phase relation (CPR) for mass supercurrent, 
$J_{\rmit{s}}(\Delta\phi)$.\cite{barone}
In the triplet case there may also exist a spin current, if 
$\Deltavec^L\nparallel\Deltavec^R$.

In non-magnetic situations it is also reasonable 
to assume a symmetry of $F_J(\Delta\phi)$ under time-reversal (TR).
If $\Delta^{L,R}$ are both ``TR invariant'', meaning
that $\Delta_ 0^{L,R}$ can be chosen real so that TR only complex
conjugates the phase factor, then
$F_J(-\Delta\phi)=F_J(\Delta\phi)$.\cite{yipgl} Similarly, the CPR then
satisfies 
$J_{\rmit{s}}(-\Delta\phi)=-J_{\rmit{s}}(\Delta\phi)$.
Let us apply these results to  the case of sufficiently {\em small
junctions}, where we can assume
$F_J(\Delta\phi)$ to be a {\em single-valued function} of $\Delta\phi$.
Making a Fourier expansion of $F_J(\Delta\phi)$ and dropping terms
based on the TR symmetry we get
\begin{equation} \label{e.expansion}
F_J(\Delta\phi) = -E_J^{(0)}-\sum_{n=1}^{\infty}E_J^{(n)}\cos(n\Delta\phi),
\end{equation}
with a similar sine expansion for $J_{\rmit{s}}(\Delta\phi)$.
However, there exist ``chiral'' states, which are \emph{not} TR invariant.
An example is the A phase of superfluid $^3$He, to which we return
shortly. In such cases the TR symmetry no longer implies
$F_J(-\Delta\phi)=F_J(\Delta\phi)$ in general,
but in special cases more general symmetries of the form 
$F_J(\beta-\Delta\phi)=F_{J}(\beta+\Delta\phi)$ may still be 
valid,\cite{yipgl} see Section 5.

An alternative expansion is to develop $F_J$ in powers of 
the order parameters
$\Delta^{L,R}$ (similar to the Ginzburg-Landau expansion). Comparing
this expansion with the Fourier expansion  (\ref{e.expansion}), we see
that the coefficient 
$E_J^{(n)}$ is proportional to $\Delta^{2n}[1+O(\Delta^2)]$, where
$\Delta$ denotes the amplitude of the order parameter.  For small
junctions the effective expansion parameter is the amplitude of the
order parameter divided by temperature,
$\Delta/k_BT_c$. Since this is small at temperatures
close to $T_c$, the lowest order terms dominate
in the series (\ref{e.expansion}). 
Usually the first order term 
$F_J^{(1)}=-E_J^{(1)}\cos(\Delta\phi)$ is most important at all $T$,
except in special cases where its amplitude is suppressed due to
symmetry reasons. These symmetries for
$^3$He or unconventional superconductors have been studied in several 
papers.\cite{viljas,viljas2,yipgl,degennes,larkin} 
All other degrees of freedom of the order parameters 
but $\Delta\phi$ (and those related to the junction itself) are now
embedded in the
coefficients $E_J^{(n)}$, and  these may be used as tuning parameters.
For example, in some cases\cite{pisquid,sfs} the sign of the (normally
positive) $E_J^{(1)}$ can be reversed, so that one obtains a ``$\pi$
junction'', where the only minimum of a sinusoidal EPR is at
$\Delta\phi=\pi$ rather than at $\Delta\phi=0$. In
$^3$He-B such a trick may be done by controlling the order parameter 
textures with magnetic fields.\cite{yippi}

For vanishing $E_J^{(1)}$ the higher order terms may still give a
finite critical current. In particular, a finite $E_J^{(2)}$ term gives
a $\pi$-periodic EPR due to the $\cos(2\Delta\phi)$ term. 
If the suppression of $E_J^{(1)}$ is only partial, then some
interesting mixtures of the $2\pi$ and $\pi$ (and shorter) periodic
components may be observed (see, for example, Ref.\ \onlinecite{tanaka}
or Ref.\ \onlinecite{viljas2} and references therein). 
In particular, it is exactly these higher-order terms of Eq.\ (\ref{e.expansion})
which result in the ``$\pi$ states'' of the $p$-wave (pinhole) junctions
discussed in Refs.\ \onlinecite{yippi,viljas2,nishida,zhang}. 
These $\pi$ states become more pronounced for $T\ll T_c$, since
the higher-order $E_J^{(n)}$'s are larger there. 
They are also the reason for the slanting of the CPR
of an $s$-wave pinhole contact at low temperatures.\cite{kulik}
However, in this case there is no way suppress $E_J^{(1)}$ in order to
single out the smaller
$\pi$ periodic components.

For {\em large junctions} we may expect two kinds of complications to
the description based on Eq.\ (\ref{e.expansion}). 
First, there is the effect of ``kinetic
inductance'': as a result of current conservation, there must be a finite
phase gradient carrying the current $J_{\rmit{s}}$ into, out of, and within the
weak link.\cite{likharev}
Thus the phase difference $\Delta\phi$ itself depends on $J_{\rmit{s}}$,
which makes $F_J(\Delta\phi)$ and $J_{\rmit{s}}(\Delta\phi)$
hysteretic (multivalued) in general. As a result, the expansion in the
cosines, Eq.\ (\ref{e.expansion}), is no longer valid as such.
The hysteresis can be modelled by introduction of the ``slanting
parameter'' and a self-consistent set of equations as in 
Refs.\ \onlinecite{likharev} and \onlinecite{avenel}.
Second, in the presence of a multicomponent order parameter
the transmission properties of the weak link (reflected by the signs
and absolute values of the coefficients $E_J^{(n)}$ above) may depend on the
order parameter configuration inside the aperture. 
This configuration may change as a function of
$\Delta\phi$, which changes the 
CPR's.\cite{monien,thuneberg88,viljas} 
Thus the order parameter must be solved
self-consistently in and around the hole, with boundary conditions
specifying $\Delta^{L,R}$ only somewhere far from the junction.
The
onset of these large-aperture effects in weak links of superfluid
$^3$He is the  main subject of the this paper. 

These changes can also be seen in the expansion of $F_J$ in the order
parameters. Instead of
$\Delta/k_BT_c$, the effective
expansion parameter in a junction of linear dimension $D$ turns out
to be $\Delta D/\hbar v_F$, the gap divided by the (ballistic) Thouless
energy.  Using standard relations the expansion parameter can also be
expressed as $D/\xi_{GL}(T)$, where $\xi_{GL}(T)$ is the Ginzburg-Landau
temperature dependent coherence length. We see that the expansion
breaks down when $D/\xi_{GL}(T)\sim 1$.  This is in agreement with our
results below, which show multivalued EPR for $D\gtrsim \xi_{GL}(T)$.

\section{SUPERFLUID $^3$He} \label{s.helium3}

Below we shall only consider the triplet $p$-wave case
where the gap vector $\Deltavec$ can be written as
$\Delta_\mu(\xvec,\kvechat)=A_{\mu i}(\xvec)\khat_i$ with a proper
choice of the spin and orbital basis functions.\cite{leggett,vollhardt}
In practice this means superfluid $^3$He, but a similar analysis may
be valid in possible triplet superconductors.
In $^3$He, in the absence of magnetic fields, there are
two stable bulk phases, the A and the B phase.
These are known to correspond to the ABM state
$A=\Delta_A \dvechat(\mvechat+\iu\nvechat)\exp(\iu\phi)$ and the BW state 
$A=\Delta_B R \exp(\iu \phi)$, respectively. Here $\Delta_{A,B}$ are
the bulk gaps of the A and B phases.
In the B phase $R(\rotaxis,\theta)$ is a rotation matrix, which may be
parametrized by the rotation axis $\omegavechat$ and rotation angle 
$\theta$. 
In the A phase $\mvechat\perp\nvechat$, and one usually defines a
third unit vector $\lvechat=\mvechat\times\nvechat$ which gives the
direction of relative orbital angular momentum of all Cooper pairs. 
Note that reversing the sign of $\dvechat$ is equivalent to a phase
shift by $\pi$, and that any phase shift by $\phi$ is equivalent to a rotation of 
$\mvechat,\nvechat$ around $\lvechat$ by angle $-\phi$. The separation of the phase factor
is therefore not unique, and it is further complicated by the
existence of textures in the $\lvechat$ field. However, in what
follows, we can assume $\lvechat$ to be constant most of the time,
and  if the same definitions are used consistently, no problems should arise.
In the B phase the gap $|\Deltavec(\kvechat)|$ is isotropic,
whereas in the A phase it has point nodes in the direction of the
vector $\lvechat$. To maximize the condensation energy, $\lvechat$ is
therefore always rigidly oriented perpendicular to solid surfaces. 
The presence of a specific direction $\lvechat$ for the Cooper pair
orbital angular momentum means
that the A phase is not time-reversal invariant. Consequently, in the
context of weak links, some of the usually ``obvious'' symmetries must
be reconsidered when A phase is involved. These symmetry properties 
are discussed below, in Section 5. 

The ABM and BW states are well-defined only in the hydrodynamic
regime, \emph{i.e.}, on large length scales. Close to surfaces, for example,
the order parameter will be modified on the coherence length scale
$\xi_0$ due to scattering of quasiparticles.
A weak link involves length scales on the order of
$\xi_0$ and a locally suppressed order parameter by definition, 
and therefore a more general treatment is
required. In what follows, we let all the components of the order
parameter vary freely close to the weak link, and the ABM or BW
forms are fixed only on boundaries in the bulk liquid.

\subsection{Symmetries and Conservation Laws} \label{s.symmetry}

The free energy of a $p$-wave spin triplet Fermi superfluid can
be expressed as a functional of the
order parameter field $A(\xvec)$. 
In a region $\Omega$ bounded by $\partial\Omega$, we assume the
free-energy expression to be of the form
\begin{equation} \label{e.geneene}
F_\Omega[A]=\int_\Omega\upd^3 xf(A,\boldsymbol\nabla A).
\end{equation}
Neglecting in this functional any terms that couple the spin and orbital degrees
of freedom or introduce preferred spin directions 
(nuclear dipole-dipole and dipole-field interactions), 
it must remain invariant under global gauge transformations [$U(1)$] and
global spin rotations [$SO(3)^s$] of $A$. These are parametrized by the independent real
parameters $\phi$ and $\theta_\alpha,\alpha=x,y,z$, respectively,
which are independent of $\xvec$. Infinitesimally
the transformations are written as
\begin{align}
A_{\mu i}&\rightarrow A_{\mu i}'=e^{\iu\delta\phi}A_{\mu i}\approx A_{\mu i}
+\iu A_{\mu i}\delta\phi \label{e.phtrans}\\
A_{\mu i}&\rightarrow A_{\mu i}'=R_{\mu\nu}(\delta\theta_\alpha)
A_{\nu i}\approx A_{\mu i}
-\epsilon_{\alpha\mu\nu}A_{\nu i}\delta\theta_{\alpha}, \label{e.rotrans}
\end{align}
where $R(\theta_\alpha)$ is a (right-handed) rotation matrix.
In this context Eq.\ (\ref{e.rotrans}) should not be considered as a \emph{passive} 
rotation of the spin-coordinate system, but rather as an \emph{active}
transformation of the physical state.
Suppose that originally the order parameter corresponds 
to a stationary point of $F_{\Omega}$, \emph{i.e.}, it is a solution of the Euler-Lagrange equations
$\delta F_\Omega/\delta A=\delta F_\Omega/\delta A^*=0$ with fixed boundary
conditions $\delta A|_{\partial \Omega}=0$, for example. 
Then for variations $\delta A$ which leave $f$ 
invariant, one finds a conservation law
$\boldsymbol\nabla\cdot\jvec_{\rmit{N}}=0$ for some generalized current
$\jvec_{\rmit{N}}$. (This is a special case of 
\emph{Noether's theorem}.\cite{byron})
In particular, for the variations of Eqs.\ (\ref{e.phtrans}) and
(\ref{e.rotrans}) the corresponding currents are the ``mass
supercurrent'' and the three
independent components of ``spin supercurrent''
\begin{align}
\jvec_{\rmit{s}}&=\frac{2m_3}{\hbar}\left(+\iu A_{\mu i}\frac{\partial f}{\partial
\boldsymbol\nabla A_{\mu i}}+c.c.\right) \label{e.massc}\\
\jvec^{\spin}_\alpha&=+\epsilon_{\alpha\mu\nu} A_{\nu i}
\frac{\partial f}{\partial\boldsymbol\nabla A_{\mu i}}+c.c., \qquad
\alpha=x,y,z. \label{e.spinc}
\end{align}
Orbital rotations do not in general keep the energy invariant and
therefore no conserved ``orbital supercurrent'' exists. 
The physical interpretation of Eqs.\ (\ref{e.massc}) and (\ref{e.spinc}) 
as ``mass'' and ``spin'' currents (and thus their normalization
constants) cannot be seen from this phenomenological 
approach, but they can be verified from microscopic theory.
Note that the
same current expressions can be obtained by inserting into 
Eq.\ (\ref{e.geneene}) the ``gauge invariant
derivative'' prescriptions 
$\delta_{\mu \nu}\partial_i \rightarrow \delta_{\mu\nu}\partial_i$
$+\delta_{\mu\nu}\iu a_i-\epsilon_{\mu\nu\alpha}b_{\alpha i}$ and
expanding to linear order in $\avec$ and $b_{\alpha i}$. 

The conservation laws $\boldsymbol\nabla\cdot\jvec_{\rmit{s}}=0$ and
$\boldsymbol\nabla\cdot\jvec^{\spin}_\alpha=0$ play an important role in the weak link
problem since the current density distributions are closely related to
the perturbations of the order parameter at the junction. Due to
the conservation of currents, the perturbations have, in principle, infinite range.
In reality the symmetry-breaking dipole interaction makes spin currents decay on
the dipole length scale $\xi_D$. 
This is much larger than the scale considered in this paper and thus
we can assume the conservation of both currents to be exact.

\subsection{Ginzburg-Landau Theory} \label{s.gl}

In numerical calculations we use the Ginzburg-Landau (GL) expansion of 
$F_{\Omega}$, which is valid at temperatures close to 
the transition temperature $T_c$, where the amplitude of $A$ is
small. The GL theory and its applications have been thoroughly discussed in various 
articles.\cite{degennes,leggett,buchholtzfetter,thuneberg87}
Taking into account the bulk and gradient terms only, the GL energy
density is
\begin{equation} \label{e.fbulk}
\begin{split}	
f(A,&\nabla A) = \\
&-\alpha\Tr(AA\!\TC)
+ \beta_{1} |\Tr(AA\!\T)|^2 \\
&+ \beta_{2} [\Tr(AA\!\TC)]^2
+ \beta_{3} \Tr(AA\!\T\!A\!\C\!A\!\TC) \\
&+ \beta_{4} \Tr(AA\!\TC\!AA\!\TC)
+ \beta_{5} \Tr(AA\!\TC\!A\!\C\!A\!\T) \\
&+ K_1\partial_{i}A_{\mu i}^*\partial_{j}A_{\mu j}
        +K_2\partial_{i}A_{\mu j}^*\partial_{i}A_{\mu j} \\
        &+K_3\partial_{i}A_{\mu j}^*\partial_{j}A_{\mu i}. 
\end{split}
\end{equation}
This includes all the linearly independent terms up to 
fourth order (second order in gradients) which are invariant
under global rotations of spin and orbital
coordinates [$SO(3)^{s,o}$], global gauge transformations [$U(1)$],
and under time reversal, \emph{i.e.}, complex conjugation. 
These include the transformations of Eqs.\  (\ref{e.phtrans}) and
(\ref{e.rotrans}).
Other terms resulting from possible magnetic 
dipole-dipole and dipole-field interactions \emph{etc.}\ could also be
included, and they would not introduce major
complications to the numerical calculation. However, these
interactions only become important on length scales $\xi_D\approx 10~\mu$m and $\xi_H$
which we assume to be much larger than the temperature-dependent
coherence length $\xi_{GL}(T)$ as defined shortly. 
This restricts the validity of our results from above on the field
and temperature scales, since $\xi_H\sim H^{-1}$ and
$\xi_{GL}(T)\sim(1-T/T_c)^{-1/2}$. In practice it may not be possible
do accurate measurements so close to the critical temperature that the
latter restriction would be a problem. Using the GL energy of 
Eq.\ (\ref{e.fbulk}) in Eqs.\ (\ref{e.massc}) and
(\ref{e.spinc}), the formulas for the currents given in
Refs.\ \onlinecite{buchholtzfetter} and \onlinecite{thuneberg87}, for example, are exactly
reproduced. 

The GL free-energy expansion was introduced above phenomenologically,
with several parameters: $\alpha,\beta_i,K_i$ and $\gamma$. 
Values for these can be calculated from quasiclassical
theory.\cite{sr,thuneberg87} All temperature-dependence of GL theory is in the coefficient
$\alpha(T)=N(0)(1-T/T_c)/3$, where $N(0)=m^*k_F/2\pi^2\hbar^2$ is the
normal-state density of states on the Fermi surface for one spin species. 
The gradient-energy parameters
are $\gamma=3$, and
$K_1/(\gamma-2)=K_2=K_3=K$$\equiv$$(7\zeta(3)/240\pi^2)$$N(0)(\hbar v_F/ k_B T_c)^2$.
These are all weak-coupling (WC) results, but the strong-coupling (SC)
corrections will not be used, since they are not very accurately
known, and are probably small. Also, $K$ and
$\alpha$ appear only in the natural length scale of GL theory, which
we use as our unit of length. This is the temperature-dependent coherence length
$\xi_{GL}(T)=\sqrt{K/\alpha}=\xi(0)(1-T/T_c)^{-1/2}$, where
$\xi(0)=\sqrt{21\zeta(3)/240\pi^2}(\hbar v_F/ k_B T_c)$ is one way to
define the zero-temperature coherence length $\xi_0$. 
The WC values for the $\beta$ parameters are 
$\beta_i^{WC}/\beta_{BCS}=(-1,2,2,2,-2)$, for
$i=1,\ldots, 5$, where
$\beta_{BCS}\equiv(7\zeta(3)/240\pi^2)N(0)/(k_B T_c)^2$.
Pressure-dependent SC corrections to these have been 
calculated in Ref.\ \onlinecite{sauls}. 
The main effect of the SC corrections is to change the 
difference $\Delta f_{AB}=f^c_A-f^c_B$ of the A and B phase condensation energies,
$f_A^c=2\alpha\Delta_A^2/2$ and $f_B^c=3\alpha\Delta_B^2/2$, 
so that the A phase can become stable at pressures above the
polycritical one, $p_0$.  The value of $p_0$ 
depends sensitively on the $\beta$ parameters, and our
fit gives it at roughly $p_0=28.7$ bar, whereas
experimentally $p_0\approx 21$ bar.  Pressures close to $p_0$ are 
important for studying weak links between the A and the B phases, since the
phase boundary must remain pinned in the aperture. For a slit of width
$W$ the condition for the stability of the boundary is 
$|\Delta f_{AB}| < 2\sigma_{AB}/W$, where $\sigma_{AB}$ is the 
surface tension of the A-B interface.\cite{viljasab} 

The boundary conditions for the order parameter 
on solid surfaces generally satisfy
$A_{\mu i}\shat_i=0$, where $\svechat$ is the surface normal.\cite{degennes}
Everywhere in this paper we use a more strict boundary condition 
$A_{\mu i}=0$. This corresponds to a microscopically rough surface,
which scatters quasiparticles in a completely diffusive way. 
Most realistic surfaces are suspected to be of this type and, in
addition, this is the simplest one to implement numerically.

\section{THE WEAK LINK PROBLEM} \label{s.link}

\begin{figure}[!tb]
\begin{center}
\includegraphics[width=0.65\linewidth]{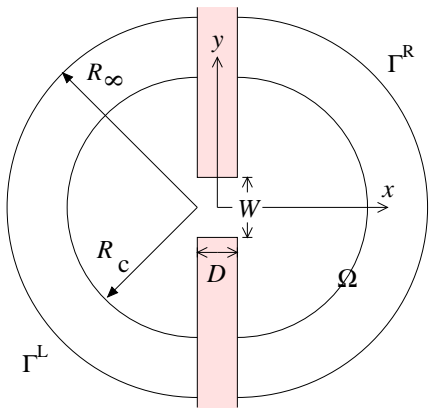}
\caption{A 2D representation of the slit-like weak link in $xy$ plane. The
circular arcs are the cutoffs used in the numerical
calculation. $R_\infty$ is a necessary artifact of the 2D
approximation, and it is on the order of the $z$
directional length $L$ of the assumed slit-type weak link.\cite{thuneberg88}
In a 3D calculation one could choose $R_\infty=\infty$ without problems. 
$R_c$ is an artificial cutoff dividing the computational effort into 
numerical (inner) and analytical asymptotic (outer) regions. In the A
phase, where the $\lvechat$ vector will be perpendicular to the wall,
the circles will be compressed to ellipsoids with the scaling 
of the perpendicular coordinate $x$ by $\sqrt{2}$.} \label{f.halfregion}
\end{center}
\end{figure}

We now apply the above considerations to describe a weak link
of the form shown in Fig.\ \ref{f.halfregion}. It is a 2D approximation
of the slit-type weak links used in experiments like those described in
Refs.\ \onlinecite{paris} and \onlinecite{avenel}. This 
``archetype'' is here considered between two infinite volumes, but 
we could also insert it between two flow channels or other restricted 
2D geometries.\cite{ullahfetter}
The origin of coordinates is placed in the middle of the junction,
with the $x$ axis running through the aperture.
Throughout, we call the sides with negative and positive $x$
coordinate the left ($L$) and right ($R$) sides, respectively.

The region $\Omega$ in Eq.\ (\ref{e.geneene}) is
now the one inside the outer circular arcs of radii $R_\infty$.
In 3D the $R_\infty$ cutoff could 
be taken to infinity. In 2D an inconvenient finite value for $R_{\infty}$ is required due to the
logarithmic $\ln(R_\infty/r)$ (and not $1/r$) asymptotic behavior of
the phase fields
[see Eqs. (\ref{e.brad}) and (\ref{e.apot}) in Appendix A]. 
This cutoff may be thought to describe some 
effective length scale, determined by the dimensions of the container,
or a distance after which the decay turns three-dimensional ($1/r$), for example. 
The $z$ directional length of the assumed slit gives a good
approximation for the latter.\cite{thuneberg88} 
In the A phase a more natural choice for the form of the 
cutoff arcs is actually ellipsoidal, with the perpendicular
coordinate scaled down by $\sqrt{(\gamma+1)/2}\approx\sqrt{2}$. This
follows from the smaller superfluid density in the direction of the orbital angular
momentum vector $\lvechat$ --- see Appendix A.

\subsection{Coupling Energy and Soft Variables} \label{s.soft}

The left and right boundary conditions are in general 
functions $A_{\mu i}^{L}(s_{L})$, $A_{\mu i}^{R}(s_{R})$ of $s_{L,R}$
which parametrize positions on the $\Gamma^{L,R}$.
The Josephson coupling energy is a functional
$F_{J}[A^{L},A^R]$$=\min_{A}F_\Omega[A]-F_0$.
Here we explicitly subtract a term $F_0$, which is the energy in
$\Omega$ for the same $A^{L,R}$, but with the aperture closed.
This is done to remove the bulk and surface contributions, but in 2D $F_J$ still
depends on $R_\infty$ due to the currents, as explained in Appendix A.
Here we have assumed that fixed boundary conditions are suitable, and that they define
the minimum uniquely. Actually, there can be several local
minima which are separated from 
each other by energy barriers. Therefore the current-phase relation
may be multivalued, each CPR branch corresponding to a
different type of minimum. Jumps (``phase slips'') between these
branches are hysteretic.

As special cases we shall consider the A and B phases of $^3$He.
There are then four main cases which can
realized for a hole in a nonmagnetic planar wall. 
Using pairs of letters to denote the $L$ and $R$ phases, respectively, these
are BB, AA with parallel ($\uparrow\uparrow$), AA with
antiparallel ($\uparrow\downarrow$) $\lvechat$'s, and the different
configurations of AB. 
The asymptotic forms in all of these cases can be treated very similarly by writing
$A_{\mu i}=R_{\mu\nu}A^{(0)}_{\nu i}(x)e^{\iu\phi}$, where the
broken-symmetry variables, or ``soft
degrees of freedom'' $R(\omegavechat,\theta)$ and $\phi$ are assumed to
be constants on $\Gamma^{L,R}$. 

The function $A^{(0)}(x)$ is the order
parameter calculated for a planar diffusive wall in the absence of a weak link,
and thus includes the suppression at the wall.
A one-dimensional minimization is required to determine it (see Appendix B).
In the B phase
$A^{(0)}=\textrm{diag}(\Delta_\perp,\Delta_\parallel,\Delta_\parallel)$,
with 
$A^{(0)}_{\nu i}(x\rightarrow\infty)=\Delta_B\delta_{\nu i}$ in bulk. 
In the A phase
$A^{(0)}_{\nu i}=a(x)\delta_{\nu x}(m^{(0)}_{i}+\iu n^{(0)}_{i})$, where
$m_i^{(0)}=\delta_{iy}$, $n_i^{(0)}=\pm\delta_{iz}$ and 
$a(x\rightarrow\infty)=\Delta_A$. Here the A and B phase bulk
gaps $\Delta_A$ and $\Delta_B$ are defined in Appendix B. 
The positive (negative) sign of $n^{(0)}_t$ chooses the
vector $\lvechat=\mvechat\times\nvechat$ to point in the
positive (negative) $x$ direction, which are two degenerate configurations.
A more familiar form for the spin part of the A phase order parameter 
is obtained by defining the $\dvechat$ vector with
$\dhat_\mu=R_{\mu\nu}\delta_{\nu x}=R_{\mu x}$.

Assuming $A^{(0)}$ to be fixed, it suffices to write $F_J$ only in
terms of the bulk order parameters, or more precisely, 
the soft degrees of freedom $\phi$ and $R$. 
If we require $F_J[A^L,A^R]$ to be invariant under global gauge
transformations and spin rotations, it should be unchanged
when $A^{L,R}$ are both multiplied by ${R^{L}}^{-1}e^{-\iu\phi^L}$.
Thus we have
$F_J=F_J[A^{(0)L}(x),{R^{L}}^{-1}R^RA^{(0)R}(x)e^{\iu\Delta\phi}]$,
where $\Delta\phi=\phi^R-\phi^L$. 
Now, since $A^{(0)L}$ and $A^{(0)R}$ are fixed by assumption, 
we see that $F_J$ can be parametrized simply as 
\begin{equation} \label{e.softcoupling}
F_J=F_J(\Delta\phi,R_{\mu i}^LR_{\mu j}^R), \quad i,j=x,y,z
\end{equation}
where an orbital-space rotation matrix $\psi_{ij}\equiv R_{\mu i}^LR_{\mu j}^R$, 
$i,j=x,y,z$ remains as an argument. In Ref.\ \onlinecite{viljas2} this
was derived only for the BB case, but we now see that it is valid more
generally.
For the A phase only the three components
$\hat d_\mu=R_{\mu x}$ out of the nine $R_{\mu i}$ are relevant. This
implies that for the AA case Eq.\ (\ref{e.softcoupling}) can be
simplified as
\begin{equation} \label{e.softcouplingAA}
F_J=F_J(\Delta\phi,\dvechat^L\cdot\dvechat^R)
\end{equation}
and for AB 
\begin{equation} \label{e.softcouplingAB}
F_J=F_J(\Delta\phi,\orbv),
\end{equation}
where we have defined an orbital-space {vector} 
$\orbc_i=\psi_{xi}=\dhat_\mu R_{\mu i}$.\cite{yipab}

Often it is also instructive to consider the following 
``tunneling'' form of the coupling energy:\cite{degennes} 
\begin{equation} \label{e.1storder}
F_J^{\rm tun}=\re[a\td A^{L*}_{\mu x}\td A^R_{\mu x} 
+ b\td A^{L*}_{\mu y}\td A^R_{\mu y}
+ c\td A^{L*}_{\mu z}\td A^R_{\mu z}].
\end{equation}
This is the lowest order term ($\propto\Delta^2$) in the
order-parameter expansion discussed in Sec.\ 2. Thus it is also the
leading term in the Fourier expansion  (\ref{e.expansion}):
$F_J^{(1)}\equiv -E_J^{(1)}\cos\Delta\phi =F_J^{\rm tun}+O(\Delta^4)$.
There are a few more assumptions coming into Eq.\ (\ref{e.1storder}).
The suppression of the order parameter near walls causes an ambiguity
what location ${\bf r}$ should be used for $A({\bf r})$ in Eq.\
(\ref{e.1storder}). Here we have used
$\td A^{L,R}$ in the bulk, and defined
$\td A = A/\Delta_{A,B}$ depending on the phase. Also, we are
assuming orthorhombic symmetry 
$[(2/m)(2/m)(2/m)]$ as for the slit above. The coefficients $a$, $b$
and $c$ are different for the different phase combinations and
temperatures, and
$c=b$ if the symmetry of the junction is cylindrical
$[(\infty/m)(2/m)]$ instead.\cite{viljas2}

\subsection{Josephson Currents} \label{s.jc}

Assume  now the situation of Fig.\ \ref{f.halfregion}, either
2D or 3D. There can be no current flow through solid walls, and 
thus
$\jvec_{\rmit{s}}\cdot\svechat=\jvec^{\spin}_\alpha\cdot\svechat=0$ on
surfaces, where $\svechat$ is the local
surface normal. Therefore, the conserved DC mass and spin currents flowing through 
the junction are given by ($\upd\Svec\cdot\rvechat>0$)
\begin{align} 
J_{\rmit{s}}&=\int_{\Gamma^R}\upd\Svec\cdot\jvec_{\rmit{s}} 
=-\int_{\Gamma^L}\upd\Svec\cdot\jvec_{\rmit{s}} 
\label{e.totmc}\\
J^{\spin}_\alpha&=\int_{\Gamma^R}\upd\Svec\cdot\jvec^{\spin}_\alpha
=-\int_{\Gamma^L}\upd\Svec\cdot\jvec^{\spin}_\alpha.
\label{e.totsc}
\end{align}
These inherit their symmetry properties from the Josephson coupling
$F_J$, and the expressions revealing this dependence
can be derived as follows. 

Consider infinitesimal variations of $F_{\Omega}[A]$
around the stationary point, which are of the form shown in Eqs.\ 
(\ref{e.phtrans}) and (\ref{e.rotrans}) but now with local variational
parameters
$\delta\phi(\xvec)$ and $\delta\theta_{\alpha}(\xvec)$. We choose them so that 
$\delta\phi|_{\Gamma^L}=\delta\theta_{\alpha}|_{\Gamma^L}=0$ and 
$\delta\phi|_{\Gamma^R}=\delta\phi^R$, 
$\delta\theta_{\alpha}|_{\Gamma^R}=\delta\theta_{\alpha}^R$. 
Due to the stationarity, only the boundary term contributes to linear
order, and using Eqs.\ (\ref{e.massc}) and (\ref{e.spinc}) we have
\begin{equation} \label{e.ech1}
\delta F_{\Omega}=(\hbar/2m_3) J_{\rmit{s}}\delta\phi^R  
-J^{\spin}_{\alpha}\delta\theta_{\alpha}^R. 
\end{equation}
On the other hand, we can similarly transform the boundary values
$\Delta\phi$ and $\psi_{ij}$, so that $\delta\Delta\phi=\delta\phi^R$
and 
$\delta\psi_{ij}=$
$-\epsilon_{\alpha\beta\gamma}R_{\beta i}^LR_{\gamma j}^R \delta\theta^R_\alpha$.
By expanding
$F_{J}(\Delta\phi+\delta\phi^R,\psi_{ij}+\delta\psi_{ij})$
$=F_J(\Delta\phi,\psi_{ij})+\delta F_J $ 
to first order in $\delta\phi^R,\delta\theta^R_\alpha$, 
and equating $\delta F_J$ with Eq.\ (\ref{e.ech1}), we find
\begin{align} 
J_{\rmit{s}}&=\frac{2m_3}{\hbar}\frac{\partial F_J}{\partial\Delta\phi} \label{e.massjc}\\
J^{\spin}_\alpha&=\epsilon_{\alpha\beta\gamma}R^L_{\beta i} R^R_{\gamma j}
\frac{\partial F_J}{\partial(R^L_{\mu i}R^R_{\mu j})},
\qquad\alpha=x,y,z. \label{e.spinjc}
\end{align}
For a BB junction these expressions were
derived in Ref.\ \onlinecite{viljas2}, but again we see that they are
more general.
Using the simplified forms of the coupling energy for AA
[Eq.\ (\ref{e.softcouplingAA})] and AB [Eq.\ (\ref{e.softcouplingAB})]
cases, Eq.\ (\ref{e.spinjc}) reduces for an AA junction to
\begin{align} 
J^{\spin}_\alpha&=[\dvechat^L\times\dvechat^R]_\alpha
\frac{\partial F_J}{\partial(\dvechat^L\cdot\dvechat^R)},
\qquad\alpha=x,y,z,
\label{e.spinjcA}
\end{align}
and for an AB junction and to  
\begin{align} 
J^{\spin}_\alpha&=\epsilon_{\alpha\beta\gamma}\dhat_\beta
R_{\gamma i}
\frac{\partial F_J}{\partial \orbc_i}, \qquad\alpha=x,y,z.
\label{e.spinjcAB}
\end{align}

Note that the expansions of $F_J$ 
leading to Eqs.\ (\ref{e.massjc}) and (\ref{e.spinjc}) work at most
locally, \emph{i.e.}, on each branch of the solution-space separately.
In some simple cases 
one might also use them in order to check the
consistency of one's numerics. 
To remove the ambiguity related to the A-phase phase factor, 
we shall \emph{always} choose $\lvechat=\pm\xvechat$ with
$\nvechat^{(0)}=\pm\zvechat$, as
explained above.
This is a natural choice since $\nvechat$-reversed 
configurations are related by time reversal. When the phase factor is
included, TR should be understood as a reversal of both $\phi$ and 
$\nvechat^{(0)}$. 
On the other hand, a reversal of $\nvechat^{(0)}$ alone is equivalent
to reversal of $\lvechat$ at constant $\phi$.

\subsection{Controlling the Boundary Conditions} \label{s.control}

Above we assumed the order parameter at the $R_\infty$ cutoff to be
fixed to the form $A(x)=RA^{(0)}(x)e^{\iu\phi}$. Here $A^{(0)}$
determines the bulk phase, including its modification at walls.
In case of the A phase, $A^{(0)}$ must always be
such that $\lvechat\parallel\pm\svechat$, where $\svechat=\pm\xvechat$ is the wall
normal, to minimize loss of condensation energy. Besides the phase angle, this leaves only the rotation
matrix $R$ to be determined by some hydrodynamic interactions. In the A phase
this reduces to fixing  $\dvechat$ with a combination of the 
dipole-dipole $\propto-(\dvechat\cdot\lvechat)^2$ and the dipole-field
$\propto(\dvechat\cdot\Hvec)^2$ bulk terms. Unfortunately, for
$\Hvec\parallel\lvechat$ the configuration remains undetermined.
Furthermore, it may be difficult to produce the most interesting situations where
$\dvechat^L\times\dvechat^R\neq 0$ by any physical means.
In principle, one way is to use magnetic fields of different
directions on the two sides. (Some more exotic ways, such as an A-B
interface, could be imagined.\cite{nishida}) 
In the B phase there is more freedom in controlling $R$.
To start with, we always assume the rotation angle $\theta$ to be in the 
minimum $\theta_0\approx 0.58\pi$ of the dipole-dipole
$\propto(\cos\theta+1/4)^2$ interaction
far from the junction.\cite{viljas2}
The remaining $\omegavechat$ vector is coupled to the wall (with normal $\svechat$)
by a surface-dipole term
$\propto-(\omegavechat\cdot\svechat)^2$, 
and to magnetic field via a bulk term $\propto-(\omegavechat\cdot\Hvec)^2$
and a surface term $\propto-(\Hvec\cdot R\svechat)^2$.
As discussed in Ref.\ \onlinecite{viljas2}, it is the surface-dipole
and the surface-field terms 
which determine $\omegavechat$ close to surfaces at low and 
high fields, respectively. For the magnetic surface configurations we
use the definitions A-D of Ref.\ \onlinecite{yippi}, but refer to
them with lowercase letters a-d. 

In addition to the rotation matrix, also the phase difference
$\Delta\phi$ must be specified. In a channel geometry this would be
replaced by specifying the phase gradient, \emph{i.e.}, the superfluid
velocity.\cite{ullahfetter} We assume that the inverse Josephson frequency
$\omega_J^{-1}=\hbar/2\Delta\mu$ is much larger than all order parameter
relaxation times (except that of $\Delta\phi$). In practice this
should be well satisfied, since $\omega_J$ tends to be no higher than
in the audio regime.\cite{berkeley} This makes
the equilibrium concepts of energy-phase $F_J(\Delta\phi)$ and
current-phase $J_{\rmit{s}}(\Delta\phi)$ relations sensible, and therefore
warrants the present calculation. However, the situation in the
A phase may again be more complicated due to the orbital viscosity 
phenomenon, which slows down the dynamics of the $\lvechat$ vector.\cite{vollhardt}
Therefore we cannot expect the calculation to properly describe 
the dynamics of phase slips, \emph{i.e.}, jumps between branches of
$J_{\rmit{s}}(\Delta\phi)$.

\subsection{About the Numerical Implementation} \label{s.implementation}

The cutoff $R_\infty$ is arbitrary, but from the 
CPR's for one choice, we may determine the CPR's
for any other one.\cite{thuneberg88} For example, we may
do the numerical calculation for $r<R_c$ where $R_c$ is small 
(see Fig.\ \ref{f.halfregion}), and then use 
Eq.\ (\ref{e.bmcphi}) or Eq.\ (\ref{e.amcphi}) to correct the phase
differences for some $r=R_\infty\gg R_c$, so that the new CPR becomes a 
``slanted'' version of the original.
Whenever spin currents are present, the spin-rotation boundary conditions
should also be modified using
Eq.\ (\ref{e.bsctheta}) or Eq.\ (\ref{e.asctheta}).
However, in the B phase this will lead to rotation matrices
which are not of the form $R(\omegavechat,\theta_0)$
that we wish to have for large $R_\infty$ (although still
$R_\infty\lesssim\xi_{D}$).

Thus the calculation should in general
be done for each (large) cutoff 
radius $R_\infty$ and each set of
spin-orbit boundary conditions separately. 
The numerical minimization may still be done inside
$R_c\ll R_\infty$, and for $r>R_c$ the asymptotic solutions of Appendix A 
are used. In our calculations, the fitting of the solutions at $R_c$ is
not done by comparing derivatives at any location on the boundary, 
but by the more physical requirement of \emph{conservation of total
mass and spin currents} over $R_c$. This leads to a somewhat cumbersome 
self-consistent iteration procedure, which is described in Appendix B
with the rest of the numerics.
Fortunately, these practical issues should not affect any of the above
or the following analysis, except through the value of $R_\infty$, and
thus a discussion of the
solutions in the asymptotic region is postponed to Appendix A. 
If these ``asymptotic corrections'' are not used, then
$R_\infty=R_c$. 


\section{CPR'S FOR AA, BB AND AB JUNCTIONS} \label{s.results}

In this section, we present a more careful analysis of the BB, AA, and
AB junctions. 
In what follows, we always present results for BB at vapor pressure $p=0$
bar, for AA at melting pressure $p=34.4$ bar, and for AB at roughly the
coexistence pressure of $p=28.7$ bar ($\Delta f_{AB}=0$).

In numerical calculations of the CPR's, the asymptotic corrections were usually
taken into account, and we used the outer cutoff
$R_\infty/\xi_{GL}\approx 30$. In this case the value for the inner
cutoff was roughly $R_c/\xi_{GL}=10$, although in principle the results
should be quite independent of it. 
As expected, the logarithmic decay of the phase corrections is slow,
and as we varied $R_c/\xi_{GL}$ and $R_\infty/\xi_{GL}$ in the range 
$10\ldots 60$, only small differences in the CPR's could be seen.
Most of the time a lattice spacing of
$\Delta x/\xi_{GL}=\Delta y/\xi_{GL}=0.5$ was used. Refinement 
did not lead to qualitative differences in the form of the CPR's, and 
only to differences of at most a few percent in the critical currents.
The minimization was usually carried out until the error (as measured
by the norm of the gradient in the $R_c$ region, see Appendix B) 
was smaller than $10^{-5}$. For given
boundary conditions, an accuracy better by more than 10 orders was
possible, but completely unnecessary, considering
the uncertainties related to the handling of the bulk cutoffs.
Current conservation at every lattice point was practically exact,
except for points on the $R_c$ cutoff, 
where only an accuracy of 
$|\boldsymbol\nabla\cdot\jvec_{\rmit{s}}|\approx 10^{-2}$ was achieved locally. 
However, the total current over the cutoff
was required to be conserved down to $10^{-4}$, if the asymptotic
corrections were used. These values refer to the $\rho_{\rmit{s}}$ and
$\rho_\perp$ units used in
Appendix A and B, which are used also in all the figures below.

\subsection{BB Junctions} \label{s.bb}

\begin{figure}[!tb]
\begin{center}
\includegraphics[width=0.99\linewidth]{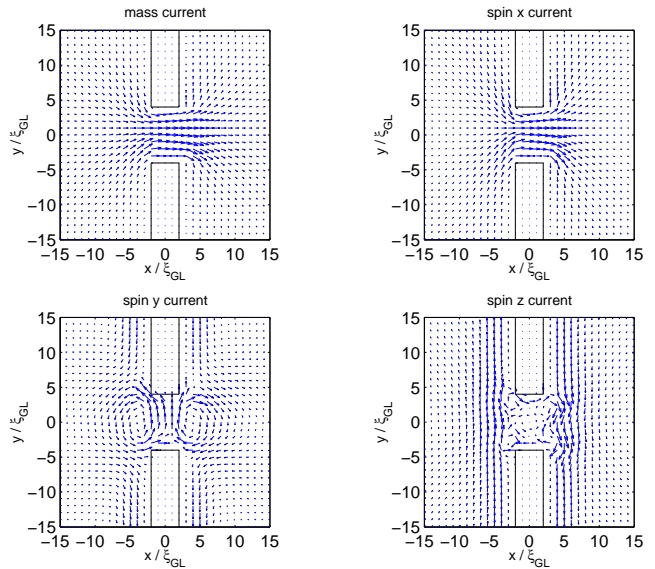}
\caption{Mass and spin current density distributions for BB with
$W/\xi_{GL}=8$, $D/\xi_{GL}=4$,
$-\omegavechat^L=\omegavechat^R=\xvechat$, and 
$\Delta\phi=0.3\pi$. The surface spin
currents of B phase are clearly visible. Close to edges, small spin
whirlpools are often formed.} \label{f.bbmsc}
\end{center} 
\end{figure}

The case of BB junctions was already considered in Ref.\ \onlinecite{viljas2}, and we
skip most of the analysis. Note that for BB time-reversal 
symmetry (complex conjugation of $A$) implies the simple relation
$F_J(-\Delta\phi)=F_J(\Delta\phi)$ in addition to $2\pi$ periodicity,
and thus $J_{\rmit{s}}(-\Delta\phi)=-J_{\rmit{s}}(\Delta\phi)$, 
$J^{\spin}_\alpha(-\Delta\phi)=J^{\spin}_\alpha(\Delta\phi)$.
Inserting Eq.\ (\ref{e.1storder}) with $\td A=R\exp(\iu\phi)$ into
Eq.\ (\ref{e.spinjc}) gives the spin current as
$J^{\spin}_\alpha=-\epsilon_{\alpha\mu\nu}[aR_{\mu x}^LR_{\nu x}^R$ 
$+bR_{\mu y}^LR_{\nu y}^R+cR_{\mu z}^LR_{\nu z}^R]\cos\Delta\phi$.
Notice that even if $F_J^{(1)}$ is identically zero, and thus the EPR
is $\pi$ periodic, the spin current may remain $2\pi$ periodic.

Despite the apparent simplicity, the BB junction has a rich spin current 
structure. This is because, in addition to the Josephson spin currents
induced by bulk boundary conditions, there exist spontaneous spin
currents close to solid surfaces.\cite{zhangwall} An example of these is
shown in Fig.\ \ref{f.bbmsc}, where small spin current loops are seen
to be stabilized in different parts of the weak link. However, these
are not ``topological'' vortices with a quantized spin velocity circulation
driving them. The cores are usually associated with a small
A-phase-like ``orbital magnetization'' 
$\propto\epsilon_{ijk}\im(A_{\mu j}^*A_{\mu k})/2$ along the $z$ axis.

\begin{figure}[!tb]
\begin{center}
\includegraphics[width=0.99\linewidth]{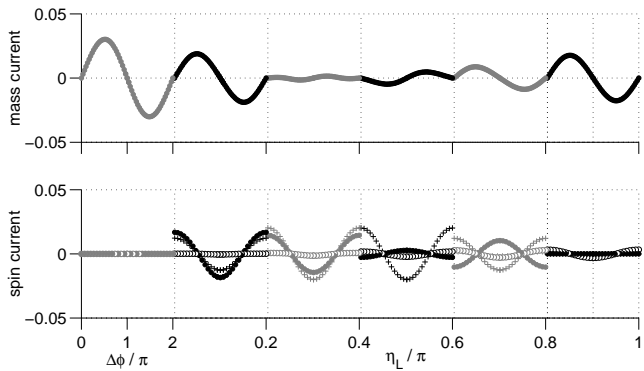}
\caption{BB current-phase relations for $W/\xi_{GL}=2$,
$D/\xi_{GL}=2$, $\omegavechat^R=\xvechat$, and $\omegavechat^L$ in $xy$ plane
with azimuthal angle $\eta_L$ with respect to $x$ axis. The spin current components are as
follows: $x$ (open circle), $y$ (plus), $z$ (closed circle).
Note that there are two scales on the bottom axis: 
a different CPR with $\Delta\phi$ in range $0\ldots2\pi$ is shown for
each of the values $\eta_L/\pi=0, 0.2, 0.4,\ldots,1.0$. 
} \label{f.bb1}
\end{center}
\end{figure}
\begin{figure}[!tb]
\begin{center}
\includegraphics[width=0.99\linewidth]{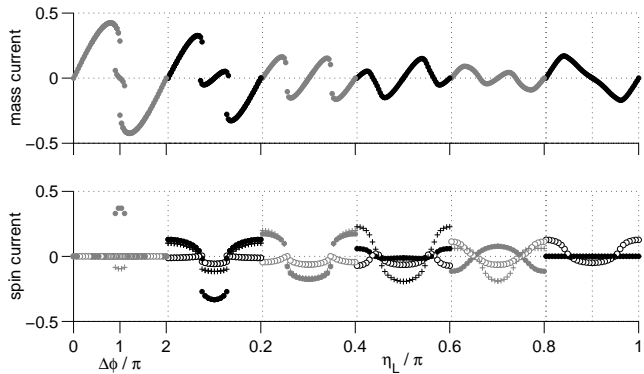}
\caption{BB current-phase relations for $W/\xi_{GL}=4$,
$D/\xi_{GL}=2$ -- otherwise same as in Fig. \ref{f.bb1}.} \label{f.bb2}
\end{center}
\end{figure}
Figures \ref{f.bb1} and \ref{f.bb2} show the CPR's for apertures of
two different widths $W$, but the same ``depth'' $D$, \emph{i.e.}, thickness of wall,
using the same series of boundary conditions: $\omegavechat^R=\xvechat$, 
and $\omegavechat^L$ is in the $xy$ plane with azimuthal angle $\eta_L$
from the positive $x$ direction. These figures are equivalent to Fig.\ 11
of Ref.\ \onlinecite{viljas2} and, while corresponding to rather
imaginary experimental conditions, they show quite
clearly the essential differences between the CPR's of narrow and wide
apertures. In the narrow case (Fig.\ \ref{f.bb1}) the behavior follows
well the perturbative picture of Eq.\ (\ref{e.expansion}): the
energy-phase and current-phase (mass or spin) relations are
non-hysteretic and close to $2\pi$ periodic (co)sine functions.
Only in cases where 
$F^{\rm tun}_J$ 
$\propto aR_{\mu x}^LR_{\mu x}^R$
$+bR_{\mu y}^LR_{\mu y}^R+cR_{\mu z}^LR_{\mu z}^R$
vanishes (as it obviously does for $\eta_L\approx0.4\pi$ and 
$\eta_L\approx0.7\pi$) a small 
$\pi$ periodic contribution due to the
remaining higher-order terms is visible, but with very small critical current. 
These higher harmonics
become stronger and produce more pronounced $\pi$ states 
at low temperatures.\cite{yippi} 
For $\eta^L\approx0.6\pi$ the CPR is sinusoidal but ``$\pi$
shifted'', whereas for $\eta^L\approx\pi$ the $\pi$ shift again
vanishes. This is different from the pinhole case of Ref.\ \onlinecite{viljas2}, where 
a $\pi$ shift is obtained for
$\omegavechat^L=\omegavechat^R=\pm\xvechat$.
This is apparently due to the difference in the symmetry of the
aperture: unlike for the pinhole, here we have $c\neq b$.

\begin{figure}[!tb]
\begin{center}
\includegraphics[width=0.99\linewidth]{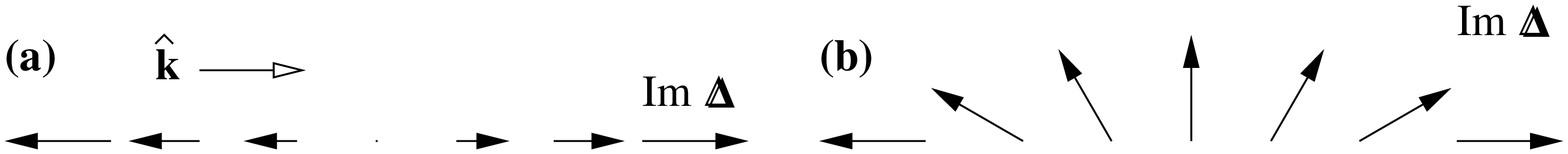}
\caption{Schematic $x$ dependence of $\im \Deltavec$ for $\kvechat=\xvechat$
at $\omegavechat^L=\omegavechat^R$ ($\theta=0$) and $\Delta\phi=\pi$ on the $0$ branch
(a) and $\pi$ branch (b). For both cases $\re \Deltavec=0$. With configuration (b) a 
lower energy may be achieved, since it avoids a singularity where the gap vanishes. 
} \label{f.escape}
\end{center}
\end{figure}
On the other hand, in
the wider case (Fig.\ \ref{f.bb2}) the CPR's are already clearly
hysteretic. The leftmost panels have
$\omegavechat^{L}=\omegavechat^{R}=\xvechat$, which is the situation studied
in Ref. \onlinecite{viljas} for the circular 3D aperture. Here the behavior
is similar: in addition to the usual ``$0$ branch'' there exists also
a ``$\pi$ branch'' around $\Delta\phi=\pi$.
Figure \ref{f.escape} depicts the corresponding behavior of the order parameter
inside the aperture at exactly $\Delta\phi=\pi$ (compare with Fig. 3
of Ref.\ \onlinecite{viljas}). Here the constant
rotation matrix has been dropped (\emph{i.e.}, $\theta=0$) so that $A$ is diagonal at the infinities.
On the ``0 branch'' the behavior of $\Deltavec(\xvec,\kvechat)$
is singular, whereas on the ``$\pi$ branch'' $\Deltavec$ 
avoids a singularity by ``escaping into the third dimension''.
The rotation of $\Deltavec$ around $\yvechat$ with growing $x$ means 
that there is a spontaneous Josephson current of $y$ directional spin flowing
through the aperture. Adding the spin rotation by $\theta=0.58\pi$ around
$\xvechat$ gives the $y$ and $z$ spin current components visible in Fig.\ \ref{f.bb2}. 
As pointed out in Ref.\ \onlinecite{viljas}, the order parameter on the
$\pi$ branch has a similar structure as in a double-core vortex,\cite{thuneberg87}
where the virtual vortex cores are inside the wall, one on each side
of the channel.

For parallel $\omegavechat$ vectors there can never be a jump from the 
0 to the $\pi$ branch, and so the conclusion in Ref.\ \onlinecite{viljas} was
that the branch may not be experimentally achievable. But this is not
so. Once the rotational symmetry of the bulk boundary conditions around
the surface normal is broken
by perturbing $\omegavechat^L$ even slightly, the hysteresis of the $0$
branch is reduced, 
and the $\pi$ branch becomes the branch of minimum
energy for $\Delta\phi\approx\pi$. 
Since $J_{\rmit{s}}'(\Delta\phi=\pi)>0$,
this ``$\pi$ state'' is even a (local or global) minimum of EPR and thus stable against small
perturbations of $\Delta\phi$. Notice how in Fig.\ \ref{f.bb2} the order parameter 
transitions between $0$ and $\pi$ minima continue to be associated
with changes in the spin currents, although a simple
interpretation as with Fig.\ \ref{f.escape} is no longer possible for
large tilting angles of $\omegavechat^L$. But for small angles
a jump onto, or away from the $\pi$ branch can be interpreted as a
\emph{phase slip by a half-quantum vortex}.\cite{viljas}
Note also that $\Delta\phi=0$ tends to remain at least a local minimum of the EPR, \emph{i.e.},
$J_{\rmit{s}}'(\Delta\phi=0)>0$, so that a sinusoidal but apparently
$\pi$ shifted EPR (``$\pi$-junction'') tends to be avoided. 
In the narrow-aperture limit $W/\xi_{GL}\lesssim 3$ the situation
$J_{\rmit{s}}'(0)<0$ appears to be more easily obtained 
(compare Figs.\ \ref{f.bb1} and \ref{f.bb2}).
The last panels of Fig.\ \ref{f.bb2} correspond to
$-\omegavechat^L=\omegavechat^R=\xvechat$, for which a $\pi$ branch with 
$J_{\rmit{s}}'(\Delta\phi=\pi)<0$ was found already in
Ref.\ \onlinecite{thuneberg88}. Our simulations confirm this old result, and
show that the $J_{\rmit{s}}'(\Delta\phi=\pi)<0$ branch of Ref.\
\onlinecite{thuneberg88} is related
to the branches with  $J_{\rmit{s}}'(\Delta\phi=\pi)>0$ (stable $\pi$ state)
by a continuous variation of the boundary conditions.

The limit between the two types of behavior, Figs.\ \ref{f.bb1} and \ref{f.bb2},
is rather clear-cut, and it only depends on the width $W$. The transition
occurs roughly at the width $W/\xi_{GL}=3$, which coincides with the
analytic critical value $\pi$ for destruction of superfluidity in an
infinite slab.\cite{fetterullah,kjaldman} However, in the weak link case the order parameter
amplitudes and critical currents always remain finite, although small.
In wide apertures the CPR's with $\pi$ states can have
relatively \emph{large critical currents} --- comparable to those of
$0$ branches in general. In the narrow apertures, however, 
there tends to be almost an order-of-magnitude difference between 
$E^{(1)}$ and $E^{(2)}$ in the GL regime.

\begin{figure}[!tb]
\begin{center}
\includegraphics[width=0.99\linewidth]{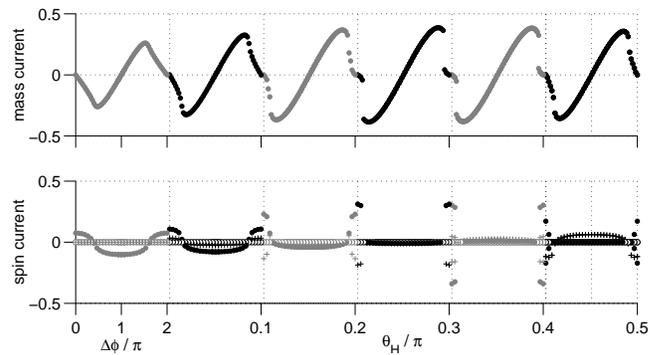}
\caption{BB current-phase relations for $W/\xi_{GL}=4$,
$D/\xi_{GL}=2$ in the magnetic ``bd'' configuration, field in the
$yz$ plane with polar angle $\theta_H=0.0,\ldots,0.5\pi$.
For $\theta_H=0$ thus
$\omegavechat^L=(-\sqrt{1/5},-\sqrt{3/5},-\sqrt{1/5})$ and
$\omegavechat^R=(-\sqrt{1/5},+\sqrt{3/5},+\sqrt{1/5})$ (see Ref.\
\onlinecite{yippi}, but note the permuted coordinate axes).
} \label{f.bb3}
\end{center}
\end{figure}

Figure \ref{f.bb3} shows the CPR's for a more physical control
parameter, namely the angle $\theta_H$ of an applied magnetic field. 
Here we must assume that the field is strong enough for the texture to be
determined by the magnetic surface interaction 
$-(\Hvec\cdot R\svechat)^2$, but small enough for $\xi_H\gg\xi_{GL}$,
since we have neglected the magnetic terms from our GL free energy.
We choose to show the ``bd'' configuration 
defined in Ref.\ \onlinecite{yippi} with $\Hvec$ in $zy$ plane, 
since this is an example of a
``$\pi$ junction'' with $W/\xi_{GL}>3$. 
In fact, a rather similar figure as Fig.\ \ref{f.bb3} exists also for the
``ad'', ``ac'' and ``bc'' configurations with field in the plane of
the wall.
For the ``ab'' or ``cd'' configurations no $\pi$ states 
nor $\pi$ shifts were seen, although many magnetic field directions
(both in-plane and off-plane) were checked.

Based on these results we conclude that a rich variety of 
$\pi$ states and $\pi$ shifts can exist in single BB apertures in the GL regime,
although it may be difficult to realize the required
$\omegavechat$ configurations experimentally. 
Approximately the behavior of Fig.\ \ref{f.bb2}, where the unreachable $\pi$ branch
becomes visible, could be
obtainable by slowly turning on $\Hvec$ while
keeping it in a suitably chosen direction. 
Without a doubt, a qualitatively similar division into small and large
aperture behavior will be valid at low temperatures also. 
To verify this, a quasiclassical calculation for the general-size
weak link would be required. However, this appears to be too demanding 
considering that there is no reason to expect any essentially new phenomena.
The critical currents and a phase diagram for BB are summarized in the next section.


\subsection{AA$\uparrow\downarrow$ and AA$\uparrow\uparrow$ Junctions}
\label{s.aa}

In AA junctions the time reversal symmetry is no longer quite so simple
as for BB, due to the fact that complex conjugation also
reverses the direction of $\lvechat$.
In both the AA$\uparrow\downarrow$ and AA$\uparrow\uparrow$
configurations TR reduces only to
$F_J(\lvechat^L,\lvechat^R,\beta-\Delta\phi)=F_J(-\lvechat^L,-\lvechat^R,\beta+\Delta\phi)$,
where $\beta=0$ or $\pi$. 
(Note again that reversing $\lvechat$ is done by flipping
the reference direction $\nvechat^{(0)}$.)
However, reversing the directions of \emph{both} $\lvechat$'s
simultaneously has no effect in our orthorhombically symmetric junction, so that 
$F_J(-\lvechat^L,-\lvechat^R,\Delta\phi)=F_J(\lvechat^L,\lvechat^R,\Delta\phi)$,
and therefore the TR symmetry reduces to that of the BB case. 
The current-phase relations therefore satisfy
$J_{\rmit{s}}(\beta-\Delta\phi)=-J_{\rmit{s}}(\beta+\Delta\phi)$ and 
$J^{\spin}_\alpha(\beta-\Delta\phi)=J^{\spin}_\alpha(\beta+\Delta\phi)$ as for BB.
Since $\pm\lvechat^{L,R}$ correspond to the same branch(es), the symmetry
also implies that the minimum of EPR is at either $\Delta\phi=0$ or $\pi$,
depending on the directions of the $\dvechat$ vectors.
In our case $\Delta\phi=0$ is the minimum if $\dvechat=\lvechat$ for both $L$
and $R$. 

For AA$\uparrow\downarrow$ the critical current will aways vanish if the aperture
has full rotation symmetry around the wall normal
$\svechat=\pm\xvechat$.
This is because an orbital rotation by $\theta$ around
$\svechat$ adds a phase $\mp\theta$ to the order parameter if
$\lvechat=\pm\svechat$, and therefore $\Delta\phi=\phi^R-\phi^L$ is changed
by $\pm2\theta$. 
Since this rotation is a symmetry operation for the aperture, it should not
change $F_J$ (\ref{e.softcouplingAA}). Thus $F_J$ should have the
periodicity 
$F_J(\Delta\phi+2\theta)=F_J(\Delta\phi)$ for \emph{all} $\theta$. This can
only be satisfied if the phase dependence of $F_J$ vanishes
altogether. In general, if the aperture has $n$-fold rotation
symmetry ($n\geq 1$), the $\uparrow\downarrow$ EPR's are $2(2\pi/n)$ periodic.
For example, for the orthorhombic slit considered 
in this paper $n=2$ and only a rotation by
$\theta=\pi$ must be a symmetry operation. This 
implies no additional restrictions on $F_J$, since it is already $2\pi$ 
periodic. The case AA$\uparrow\uparrow$ has nontrivial $2\pi$ periodic 
$F_J$'s irrespective of the geometry, except for the special 
case $\dvechat^L\cdot\dvechat^R=0$ (see below). 

\begin{figure}[!tb]
\begin{center}
\includegraphics[width=0.99\linewidth]{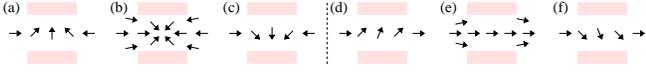}
\caption{Different configurations for the $\lvechat$ field in an
aperture. (a-c) are for antiparallel and (d-f) are for parallel
$\lvechat$'s. } \label{f.lconfs}
\end{center}
\end{figure}
\begin{figure}[!tb]
  \begin{minipage}[t]{.49\linewidth}
     \centering
     \centering \includegraphics[width=0.95\linewidth]{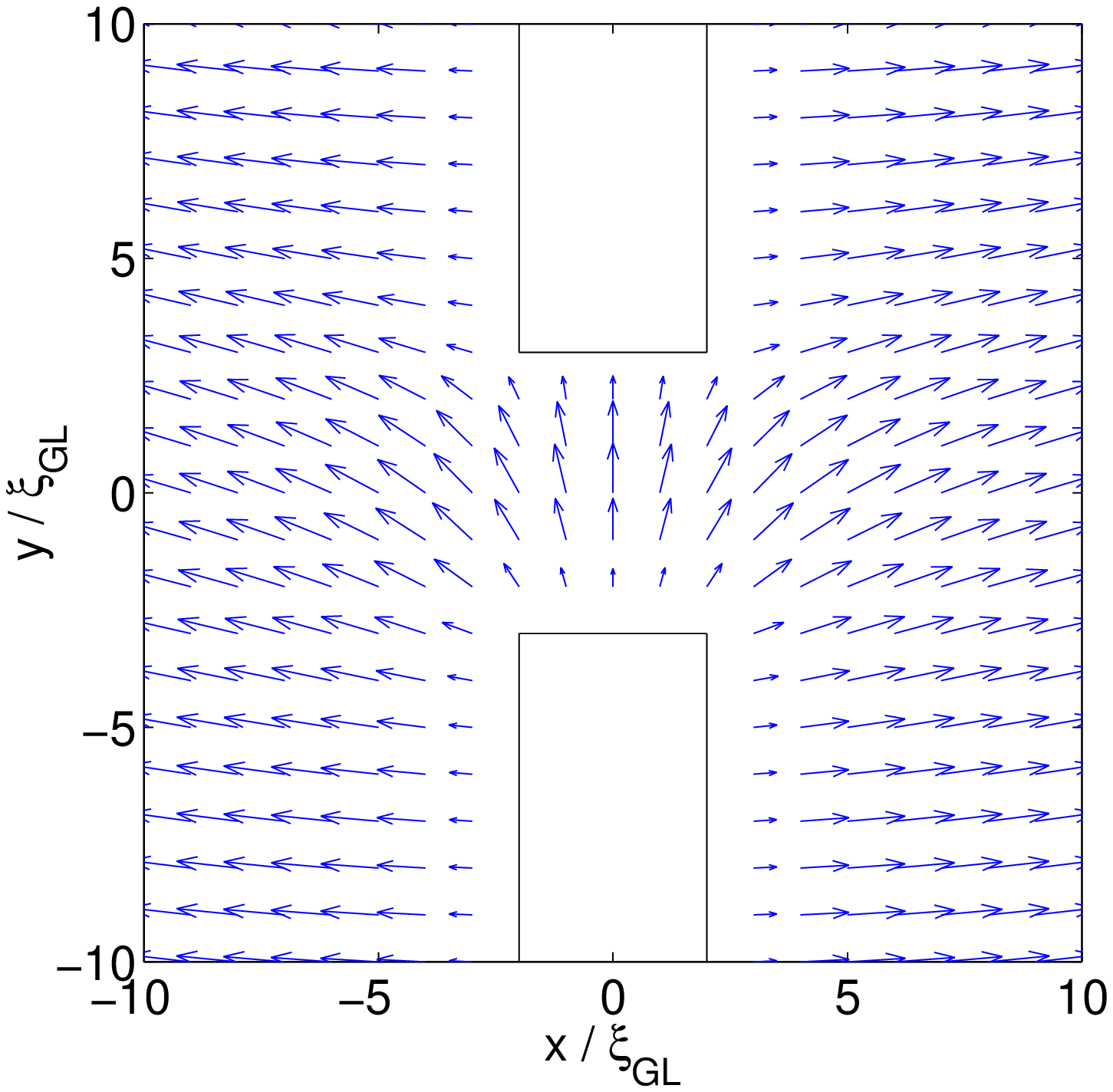}
  \end{minipage}
\hfill
  \begin{minipage}[t]{.49\linewidth}
    \centering \includegraphics[width=0.95\linewidth]{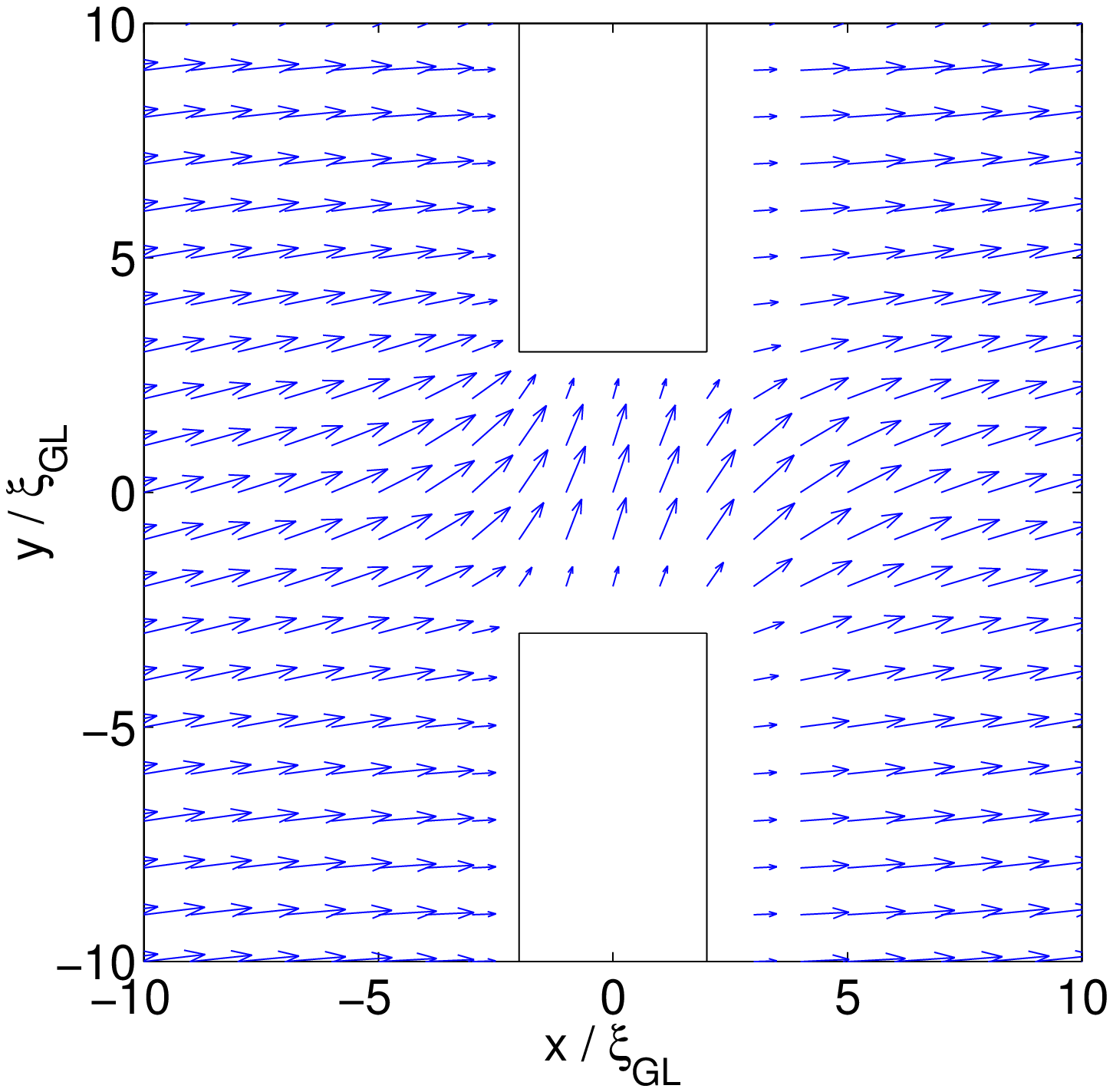}
  \end{minipage}
  \begin{minipage}[t]{.49\linewidth}
     \centering
     \centering \includegraphics[width=0.95\linewidth]{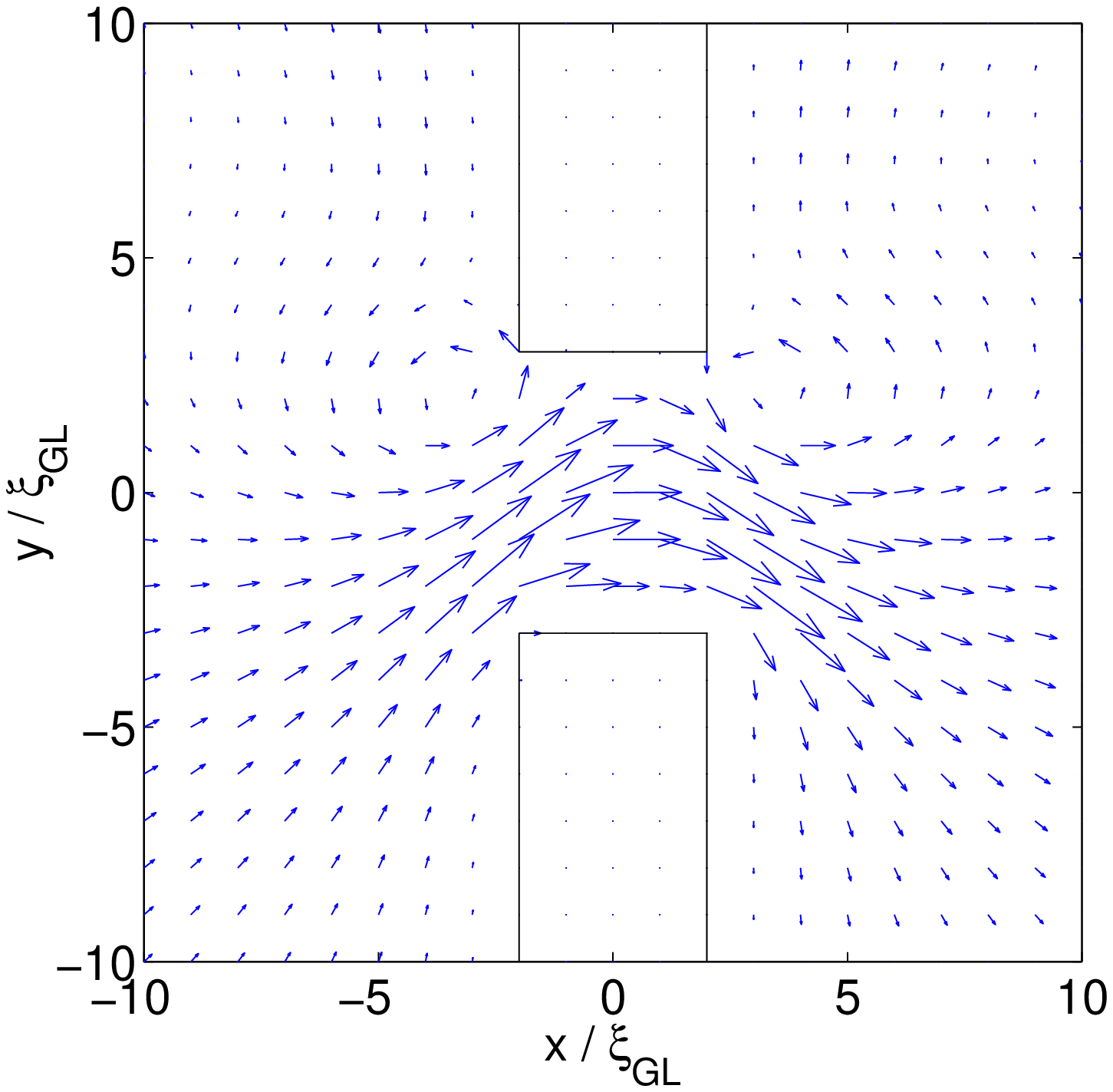}
  \end{minipage}
\hfill
  \begin{minipage}[t]{.49\linewidth}
    \centering \includegraphics[width=0.95\linewidth]{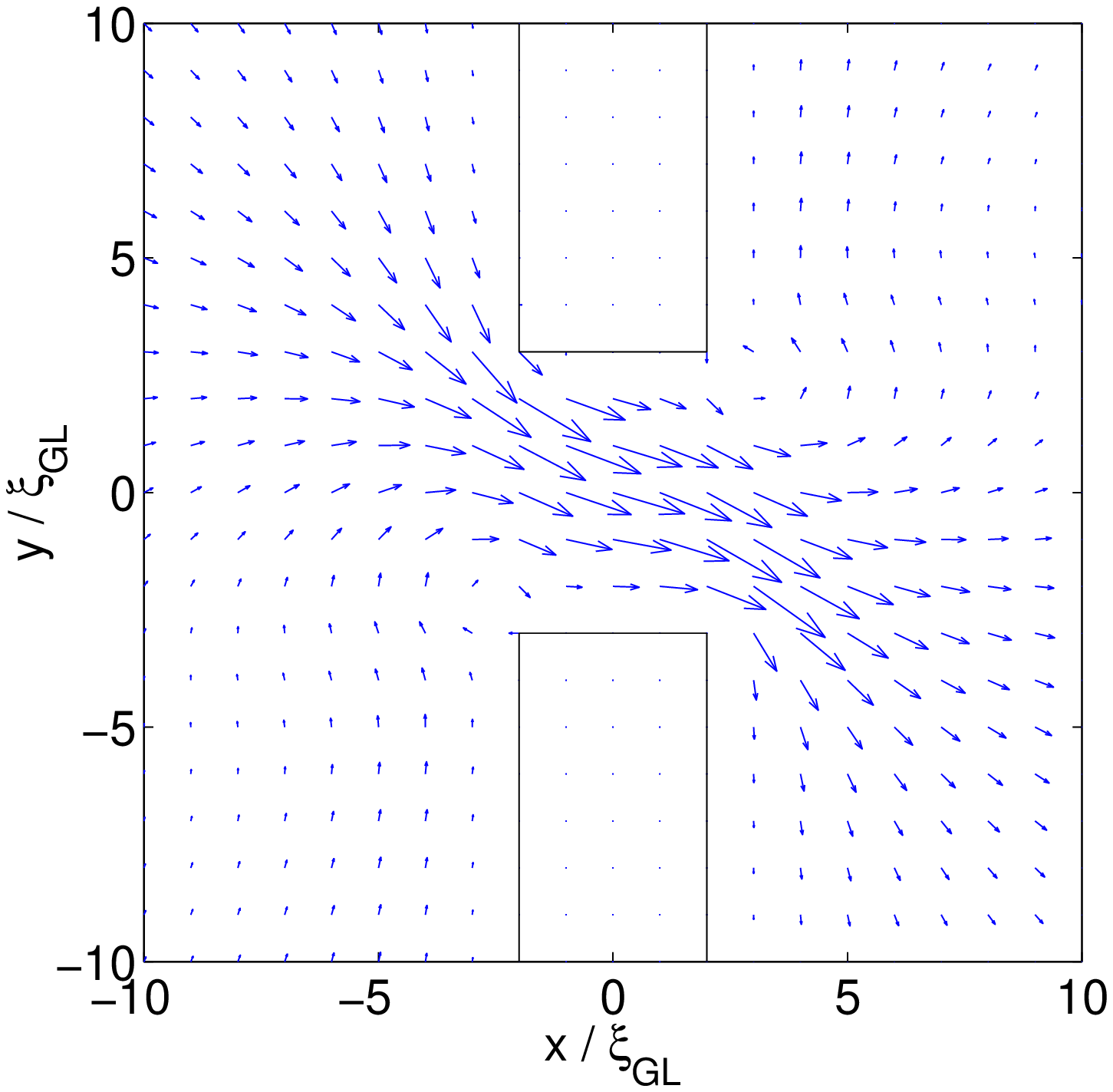}
  \end{minipage}
\caption{The top panels show the $\lvechat$ texture for $W/\xi_{GL}=6$,
$D/\xi_{GL}=4$, $\Delta\phi=0.3\pi$ for $AA\uparrow\downarrow$ (left) and the same for
$AA\uparrow\uparrow$ (right). The corresponding mass currents are shown in
the bottom panels.} \label{f.lfields} 
\end{figure}
Figure \ref{f.lconfs} represents the different configurations
for $\lvechat$ inside the aperture. The middle one for
both $\uparrow\downarrow$ (b) and $\uparrow\uparrow$ (e) is symmetrical with
respect to the $x$ axis. Case (b) is very singular and does not usually
correspond to the minimum energy in large apertures, but may be
metastable at least for $\Delta\phi\approx0$. Here a radial singularity is depicted, but a
``hyperbolic'' one is also possible. In some cases $\lvechat$ can also
escape from the plane. The $\uparrow\uparrow$ configuration (e), on
the other hand, is usually the minimum-energy configuration for
$W\gg D$ and $\Delta\phi\approx 0$, as may be expected. 
The other configurations correspond to
pairs of degenerate cases, where $\lvechat$ bends asymmetrically, as shown in
Fig.\ \ref{f.lfields} in more detail. Actually, the quantity shown is
$\epsilon_{ijk}\im(\td A_{\mu j}^*\td A_{\mu k})/2$, 
where $\td A=A/\Delta_{A}$, which reduces to $\lhat_i$ in the bulk.
Figure \ref{f.lfields} also shows the mass current density fields
associated with a phase difference $\Delta\phi=0.3\pi$. Because the
superfluid density is twice as large in directions perpendicular to 
$\lvechat$ than for those in parallel with it, the bending of the
$\lvechat$ fields results in asymmetrical current distributions 
close to the junction. 
These effects result in a more 
complicated numerical problem than for the isotropic B phase (see
Appendix A).

\begin{figure}[!tb]
\begin{center}
\includegraphics[width=0.99\linewidth]{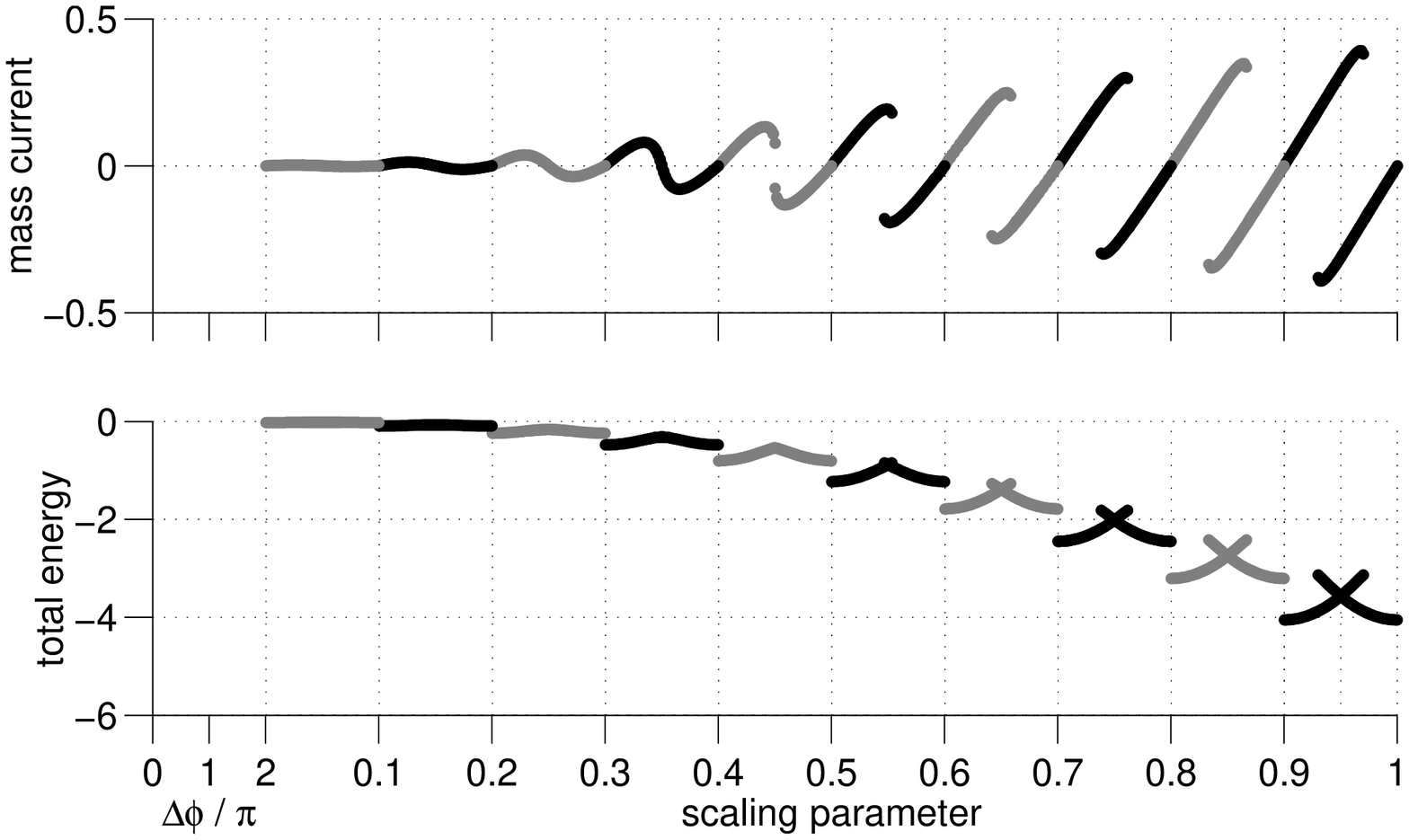}
\caption{AA$\uparrow\downarrow$ current-phase relations for aspect
ratio $W:D=2:1$, $\lvechat^L=\dvechat^L=-\xvechat$, 
$\lvechat^R=\dvechat^R=\xvechat$, as a function
of the scaling parameter $l=D/(4\xi_{GL})$.} \label{f.aanpscaling}
\end{center}
\end{figure}
\begin{figure}[!tb]
\begin{center}
\includegraphics[width=0.99\linewidth]{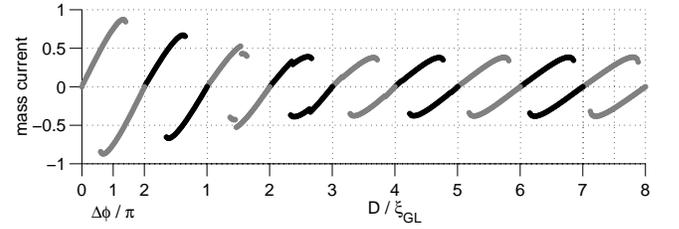}
\caption{AA$\uparrow\uparrow$ CPR's for 
$W/\xi_{GL}=6$ and $\dvechat^{L,R}=\lvechat^{L,R}=\xvechat$ 
as a function of $D/\xi_{GL}$. For small $D$, the symmetric
$\lvechat$ texture is the minimum-energy configuration, whereas
for large $D$ it is one of the asymmetrical ones. In the range 
$D/\xi_{GL}=2\ldots4$ the configuration depends on $\Delta\phi$,
which makes the CPR discontinuous. Note that for $D/\xi_{GL}\gtrsim 3$ 
the critical current no longer changes, but the CPR becomes slowly
more hysteretic. Compare with Fig. \ref{f.icall} below.} \label{f.aappd}
\end{center}
\end{figure}
Figure \ref{f.aanpscaling} shows
CPR's for $\uparrow\downarrow$ and an aperture of fixed aspect ratio,
but different dimensions. 
No $\pi$ branches exist as for BB, and the 
corresponding figure for $\uparrow\uparrow$ would be very similar.
It was noticed that, both in  $\uparrow\downarrow$ and
$\uparrow\uparrow$ cases, when a jump to
another branch of $J_{\rmit{s}}(\Delta\phi)$ occurred, the asymmetric texture shifted from one 
degenerate configuration (a or d) in Fig.\ \ref{f.lconfs} to the other
(c or f). At least in the $\uparrow\downarrow$ case this could be understood
by remembering that a phase slip of $\Delta\phi$ 
by $2\pi$ is equivalent to a rotation of the orbital space around
$\svechat$ by $\pi$. 
However, these results are due to the 
minimization algorithm, and do not necessarily 
correspond to physical dynamics.

As pointed out above, for AA$\uparrow\uparrow$ at small $D$ and
$W/\xi_{GL}\gtrsim 3$
the minimum-energy configuration is that of Fig.\ \ref{f.lconfs}e 
when $\Delta\phi=0$ and no current flows. But when
$\Delta\phi$ is increased, a change to one of the asymmetric $\lvechat$
configurations (d or f) will occur, which causes a sudden reduction of the
current and thus a discontinuity in CPR. This is illustrated in
Fig.\ \ref{f.aappd} --- see also Figs.\ \ref{f.icall} and 
\ref{f.pdiag} below. The larger $D$ is, the smaller is the
critical $\Delta\phi$ for this event. An accurate calculation of this
critical value is difficult, however, since it depends
on whether or not the asymptotic correction for $\lvechat$ is used, and
on how far the $R_c$ cutoff is taken.
This introduces some amount of
additional uncertainty in the forms of the CPR's.
For AA$\uparrow\downarrow$ one of the asymmetric configurations (a or c) 
is always the minimum one (except for small $D$ and $W$, see Fig.\ \ref{f.pdiag})
and thus in Fig.\ \ref{f.aanpscaling} all the CPR's are smooth and well-behaved.

For $\dvechat^L\cdot\dvechat^R=0$ the CPR's are $\pi$ periodic in both
AA configurations, as can be expected by noticing that $F^{\rm tun}_J$ 
[Eq.\ (\ref{e.1storder})] vanishes in this case.
However, as discussed in Section 4, there appears to be no easy physical
way in which a situation with $\dvechat^L\times\dvechat^R\neq0$ could be 
achieved. 
According to Eq.\ (\ref{e.spinjcA}) 
this more-or-less rules out spin currents, and thus seems to render
the AA CPR's rather un-interesting, apart from the effects of
Fig.\ \ref{f.aappd} arising from the $\lvechat$ texture. 
Theoretically this is not a problem, of course, and
if we force the configuration $\dvechat^L\cdot\dvechat^R=0$, 
then the following observations can
be made. 
In the small-hole limit (CPR's continuous) the critical currents are
considerably suppressed, and $J_{\rmit{s}}(\Delta\phi) \approx J_c\sin(2\Delta\phi)$.
However, in the large-hole limit (CPR's hysteretic), the critical
currents are not significantly affected. The CPR's look just like in
Figs.\ \ref{f.aanpscaling} or \ref{f.aappd}, but with the $\pi$ branches
being added in between each $0$ branch.
As the $\dvechat$'s are gradually changed from $\dvechat^L\cdot\dvechat^R=0$,
either the $0$ or the $\pi$ branches begin to rise in energy, and the
corresponding CPR branches get smaller (similar to metastable ``$\pi$
states'' in BB). Finally the branches become
unreachable and $2\pi$ periodicity without any $\pi$ states is regained.
Again, these observations hold for both AA$\uparrow\downarrow$ and
AA$\uparrow\uparrow$ alike. The most notable difference between them
is in the critical currents, which are summarized in the next
section.

\subsection{AB Junctions} \label{s.ab}

The case of an AB junction is the most complicated one to analyze.
In this case the requirement of time reversal symmetry implies
$F_J(\lvechat,\beta-\Delta\phi)=F_J(-\lvechat,\beta+\Delta\phi)$, where
$\beta=0$ or $\pi$ as in AA. 
This no longer simplifies to the intuitive BB form as it did for our AA junctions.
Rather the EPR's for $\lvechat=\pm\xvechat$ form completely separate,
unsymmetrical branches, which are $2\pi$ periodic in general, and
which can both still separate into
hysteretic sub-branches as usual. The minimum of the EPR is also not
restricted to $\Delta\phi=0$ or $\pi$ on either branch (see
Fig.\ \ref{f.ab1} below). Similarly, the
currents only satisfy relations between the $\pm\lvechat$ branches
$J_{\rmit{s}}(\lvechat,\beta-\Delta\phi)$
$=-J_{\rmit{s}}(-\lvechat,\beta+\Delta\phi)$ and 
$J^{\spin}_\alpha(\lvechat,\beta-\Delta\phi)=J^{\spin}_\alpha(-\lvechat,\beta+\Delta\phi)$.
The roles of $\Delta\phi=0$ and $\pi$ can again be interchanged by
flipping the $\dvechat$ of the A phase.

To analyze the AB case further, we consider the tunneling 
term, Eq.\ (\ref{e.1storder}). Assuming A phase on the left
and B phase on the right, we again introduce the
orbital space vector $\orbc_i=\dhat_\mu R_{\mu i}$ and find that
$F_J^{\rm tun}=\re\{[b\orbc_y(\mhat_y\pm\iu\nhat_y)$
$+c\orbc_z(\mhat_z\pm\iu\nhat_z)]\exp(\iu\Delta\phi)\}$.
We see that if $\dhat_\mu=\pm R_{\mu i}\shat_i$, \emph{i.e.}, $\orbv\parallel\svechat$, then
$\orbv\perp\mvechat,\nvechat$ and $F_J^{\rm tun}$ vanishes.
This is true, for example, in the most symmetric (and, in the absence
of magnetic fields, most probable) 
situation where both $\dvechat$ and the B phase rotation axis 
$\omegavechat$ are
perpendicular to the wall, but not necessarily if either of them 
deviates from this configuration. 
Incidentally, $\dhat_\mu=\pm R_{\mu i}\shat_i$ is also the
minimum-energy configuration for a free, planar AB interface, when
$\svechat$ is the interface normal.\cite{cross77}

More general conclusions can be based on the general form
$F_J$ (\ref{e.softcouplingAB}) in the symmetric case 
$\orbv\parallel\svechat$.  Similarly
as in the case of AA$\uparrow\downarrow$, a rotation of  the orbital
space around $\svechat$ by angle $\theta$ is now equivalent to a shift
of the phase difference $\phi$ by $\pm\theta$. (The factor is different
here since the B phase side is not affected by the rotation.)
In the case of a circularly symmetrical aperture we must then have
$F_J(\Delta\phi+\theta)=F_J(\Delta\phi)$ for all $\theta$ which, again, means that the phase
dependence must vanish to all orders. For the slit junction this implies only 
$F_J(\Delta\phi+\pi)=F_J(\Delta\phi)$, \emph{i.e.}, that the energy-phase 
relations are $\pi$ periodic. 
Generally, if the aperture has $n$-fold
rotation symmetry, then the CPR's must be $2\pi/n$
periodic. However, these simple
conclusions no longer hold when $\orbv\nparallel\svechat$, 
and in general the nontrivial $2\pi$ periodic behavior is obtained 
in any aperture geometry.

\begin{figure}[!tb]
\begin{center}
\includegraphics[width=0.99\linewidth]{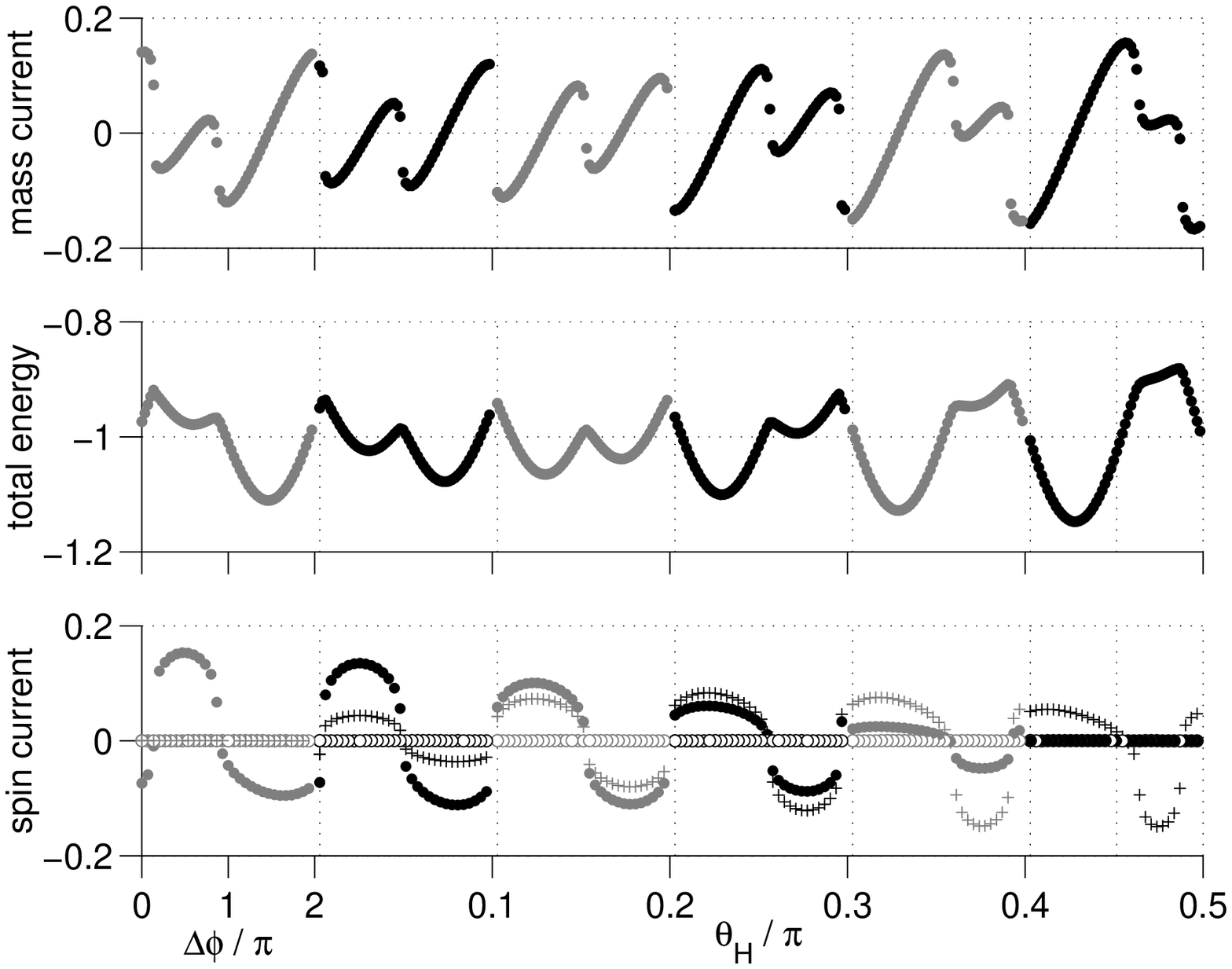}
\caption{AB current-phase relations for $\lvechat^L=\xvechat$, $W/\xi_{GL}=4$,
$D/\xi_{GL}=2$ and B phase in the magnetic ``a''
configuration,\cite{yippi} 
with field in the $yz$ plane with polar angle $\theta_H$.
Thus $\dvechat=\lvechat$ for all $\theta_H$, and, for example, the B
phase rotation axes for $\theta_H=0$ and $\theta_H=0.5\pi$ are
$\omegavechat=(+\sqrt{1/5},-\sqrt{3/5},+\sqrt{1/5})$ and
$\omegavechat=(+\sqrt{1/5},+\sqrt{1/5},+\sqrt{3/5})$,
respectively. The EPR's and spin CPR's for $\lvechat^L=-\xvechat$
branch (not shown)
could be obtained by mirroring with respect to $\Delta\phi=\pi$, and the
mass CPR's with the mirroring and a change of sign (see text).} \label{f.ab1}
\end{center}
\end{figure}
\begin{figure}[!tb]
\begin{center}
\includegraphics[width=0.99\linewidth]{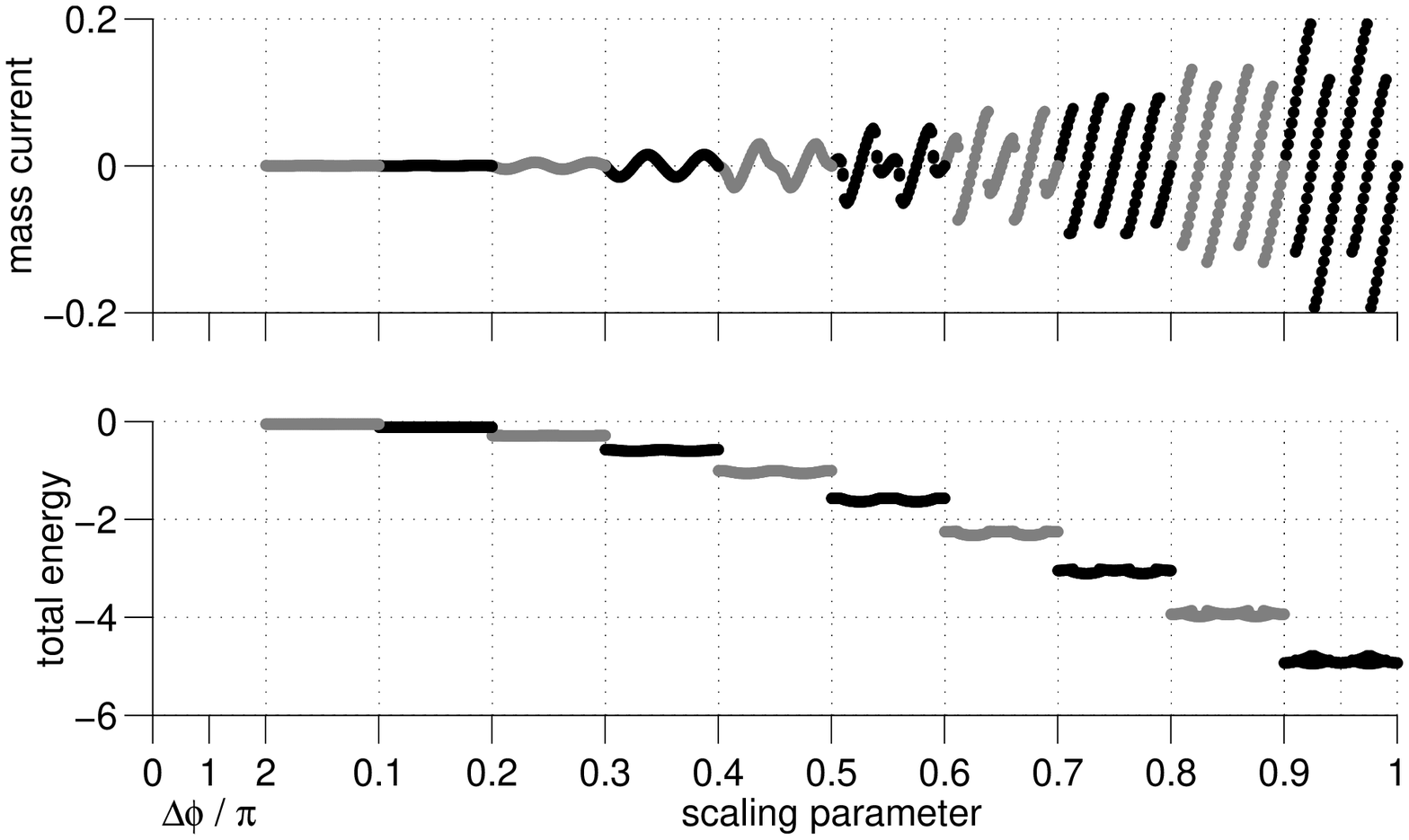}
\caption{AB current-phase relations for aspect
ratio $W:D=2:1$, and $\lvechat^L=\dvechat^L=+\xvechat$, 
$\omegavechat^R=\xvechat$ (or $\dhat_\mu=\pm R_{\mu i}\shat_i$ in general) 
as a function of the scaling parameter $l=D/(4\xi_{GL})$.} \label{f.abscaling}
\end{center}
\end{figure}

Figure \ref{f.ab1} shows examples of CPR's when $\dvechat$ and
$\omegavechat$ vectors are controlled by a field $\Hvec$ strong enough so that
$\dvechat$ points in the direction corresponding to minimum possible dipole energy
$-(\dvechat\cdot\lvechat)^2$ allowed by a strict condition $\dvechat\cdot\Hvec=0$.
Here $\omegavechat$ is chosen to be in the ``a'' configuration.\cite{yippi}
The failure of the phase-inversion symmetry 
$F_J(-\Delta\phi)=F_J(\Delta\phi)$ is clearly visible; Figure \ref{f.ab1} is for
$\lvechat=\xvechat$, and the branch for $\lvechat=-\xvechat$ is
obtained by using the above symmetries for $F_J$. Figure
\ref{f.abscaling}, on the other hand, shows the CPR's for a fixed
aspect ratio $2:1$ (as in Fig.\ref{f.aanpscaling} for
AA$\uparrow\downarrow$) but different dimensions in the
$\orbv\parallel\pm\svechat$ case.
The CPR's are $\pi$ periodic as they should, and even show the presence of
strong $\pi/2$-periodic admixtures, or separate ``$\pi/2$
branches''.

\begin{figure}[!tb]
  \begin{minipage}[t]{.6\linewidth}
    \centering \includegraphics[width=0.99\linewidth]{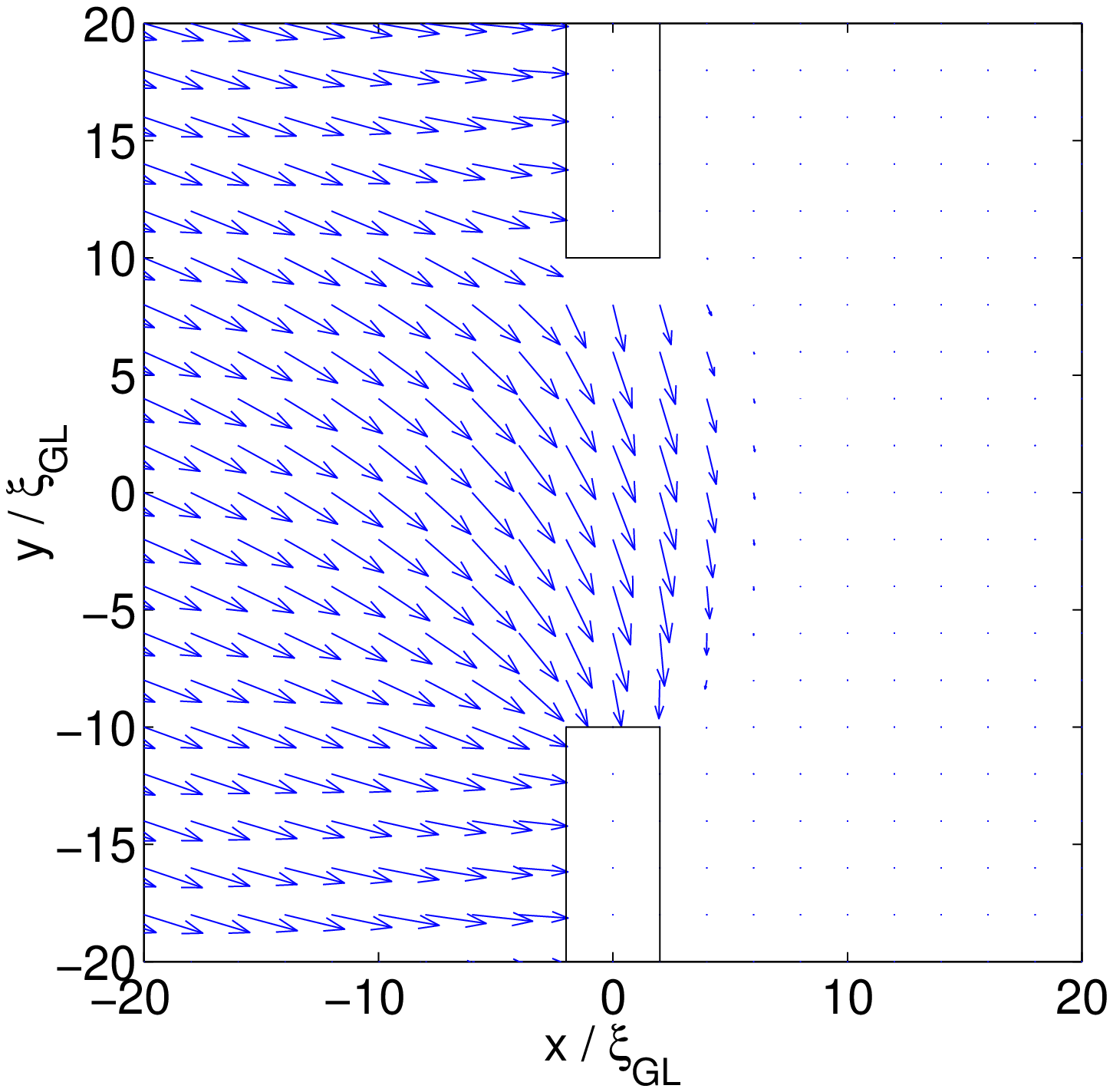}
  \end{minipage}
\hfill
  \begin{minipage}[t]{.7\linewidth}
     \centering
     \centering \includegraphics[width=0.99\linewidth]{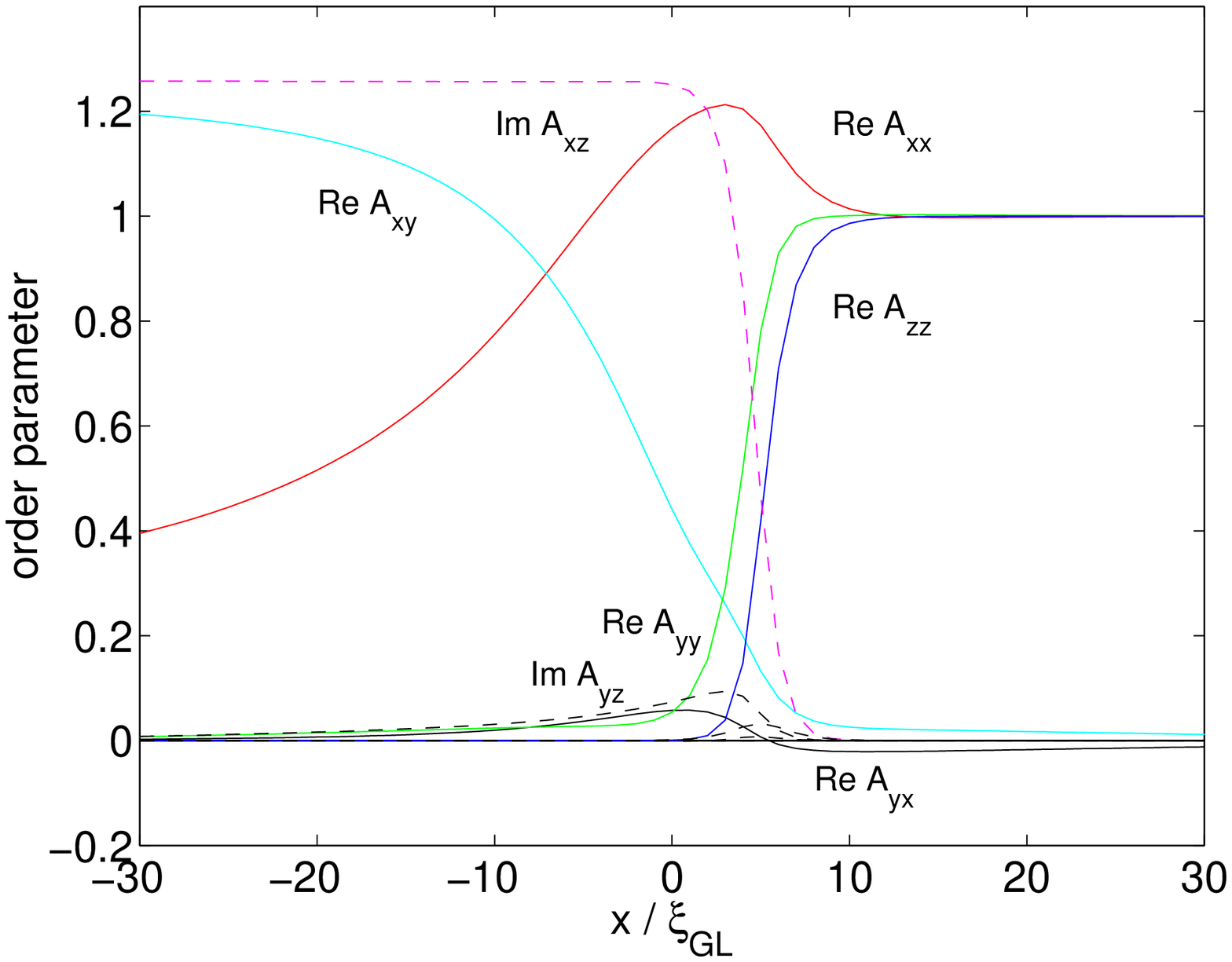}
  \end{minipage}
\caption{The $\lvechat$ field terminating at the pinned A-B interface
for $W/\xi_{GL}=20$, $D/\xi_{GL}=4$ at $p=34.4$ bar.
Also shown is a slice (at $y=0$) of the order parameter through the interface.
Here $\Delta\phi=0$ and $\dhat_\mu=R_{\mu i}\shat_i$, where the
constant $R$ matrix has been dropped in both phases for clarity.
The cut-off is at $R_c/\xi_{GL}=30$, \emph{i.e.}, at $x/\xi_{GL}=-30/\sqrt{2}$ in
the A phase. The asymptotic corrections make the order parameter
transition at the cutoff smooth.} \label{f.ablfield}
\end{figure}
The order parameter profile on the $x$ axis which goes through the AB
interface, and the form of the $\lvechat$ field there are shown in
Fig.\ \ref{f.ablfield}. Here we have $\Delta f_{AB}>0$ ($p>p_0$) for
illustration purposes, although the CPR's were calculated for $\Delta_{AB}=0$.
The boundary of the $\lvechat$ field, \emph{i.e.}, the
A-B interface is thus seen to bulge into the B phase. 
Even at $\Delta f_{AB}=0$ the interface always settles into the
B-phase end of the channel.\cite{viljasab}
As in the AA cases, $\lvechat$
tends to bend parallel to $\yvechat$ inside the hole. This is the
preferred configuration also for the A-B interface itself.
The order parameter components shown on the right are the same as in
Ref.\ \onlinecite{schopohl} or Ref.\ \onlinecite{thuneberg91}, except for the
components arising from the bending
of $\lvechat$ for $x\rightarrow-\infty$.
Similarly to the AA cases, a phase slip (jump from branch to another) 
was usually found to be associated
with a transition of the $\lvechat$ texture. Textures with
mirror-symmetry with respect to the $xz$ plane and $\lvechat$ pointing
out of the plane in the $\pm\zvechat$ direction at the interface were
also often seen on some branches. 


\section{SUMMARY AND PHASE DIAGRAMS} \label{s.summary}

\begin{figure}[!tb]
\begin{center}
\includegraphics[width=0.99\linewidth]{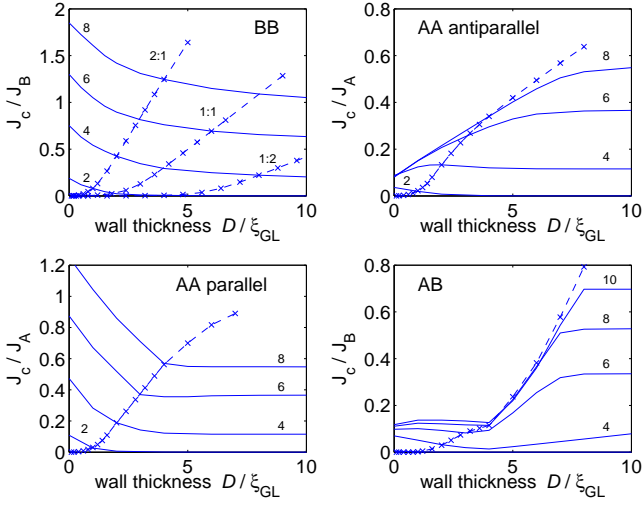}
\caption{Critical currents for the four main phase combinations. 
The solid curves correspond to different widths $W/\xi_{GL}$,
indicated by the numbering. BB is the same as in
Ref. \onlinecite{ullahfetter}, but units differ by a factor of 10. 
The dashed lines show $J_c$ for
apertures of fixed aspect ratio $W:D=2:1$ --- for BB two others are also shown. 
The asymptotic behavior at small sizes is the same for both AA cases.
For AA$\uparrow\downarrow$, AA$\uparrow\uparrow$, and  AB
the $J_c$'s become equal at large $D$. All $J_c$'s are calculated for
the most symmetric boundary conditions only. For other choices, the values
(at small $D$) will tend to be smaller in BB and AA junctions, but larger in
the AB junctions. The sharp features result from transitions of
$\lvechat$. They are explained in the text and in Fig.\ \ref{f.pdiag}. 
}
\label{f.icall}
\end{center}
\end{figure}
\begin{figure}[!tb]
\begin{center}
\includegraphics[width=0.9\linewidth]{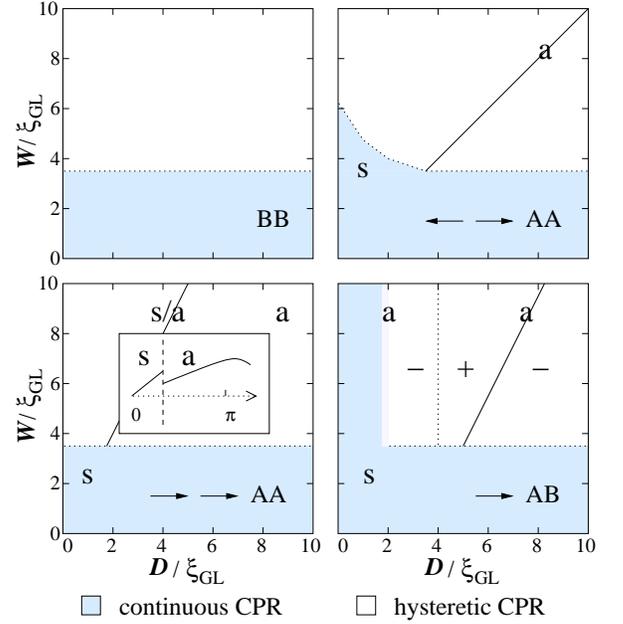}
\caption{Phase diagram for the 2D BB, AA, and AB junctions.
For $W/\xi_{GL}\lesssim 3$ the order parameter is strongly suppressed
inside the hole, and the CPR's are continuous. For
AA$\uparrow\downarrow$ this region is slightly larger. 
The letter 's' denotes regions where the A phase $\lvechat$ field is
symmetric, and 'a' regions where it is antisymmetric; 's/a' means that
the configuration depends on $\Delta\phi$. 
The sharp features in Fig.\ \ref{f.icall} are always associated with a change in the CPR
branch which determines $J_c$. In AA$\uparrow\uparrow$, the branch with 's'
configuration, which exists for small $\Delta\phi$, determines
$J_c$ to the left of the solid line --- the associated type of CPR is sketched in
the inset. To the right of the line, $J_c$ becomes essentially independent of
$D$, and the weak link approaches the ``infinite slab'' limit.\cite{fetterullah} 
In AB, the regions marked by $+$ or $-$ denote whether $J_c$ arises from the 
$0$ ($+$) or the $\pi/2$ ($-$) branch of the $\pi$ periodic
CPR's (cf. Fig.\ \ref{f.abscaling}). To the right of the second
transition marked by the solid line, the infinite slab limit is again reached.
In AA$\uparrow\downarrow$ no sharp branch transition
exists, but the solid line has roughly the same interpretation. 
}
\label{f.pdiag}
\end{center}
\end{figure}

Figure \ref{f.icall} summarizes the critical mass currents $J_c$ for BB, AA, and
AB junctions, plotted in the units defined in Appendix B. 
They have been calculated for several sizes in the range
$W/\xi_{GL},D/\xi_{GL}=0\ldots10$, which is certainly the most
relevant one. For essentially larger dimensions the
CPR's become strongly multivalued, and the aperture cannot be
considered as a weak link. Figure \ref{f.pdiag}, on the other hand, is a phase
diagram which relates the changeover between
continuous and hysteretic CPR's, and some of the $\lvechat$ texture 
transitions to well-defined regions of $D,W$ plane.
The results were calculated using a lattice spacing of 
$\Delta x/\xi_{GL}=\Delta y/\xi_{GL}=0.5$, and with a cutoff $R_c/\xi_{GL}=20$, but
with no asymptotic corrections. In AA and AB, some amount of
error will necessarily remain close to regions where $\lvechat$ transitions
occur, since $\lvechat$ is not allowed to vary freely beyond $R_c$,
which makes the texture somewhat too rigid. 
All of the results of Figs.\ \ref{f.icall} and \ref{f.pdiag} 
are for the most symmetric configurations only: in BB 
$\omegavechat^L=\omegavechat^R$, in AA $\dvechat^L=\pm\dvechat^R$, and
in AB $\dhat_\mu=R_{\mu i}\shat_i$. However, at least the narrow-aperture
limit $W/\xi_{GL}\lesssim 3$ is independent of any the boundary
conditions. The regime with $W/\xi_{GL}\lesssim 3$ (for small $D$) is 
where the description due to Eq.\ (\ref{e.expansion}) is supposed to
be valid, whereas for larger
apertures hysteresis begins to set in. This transition in behavior is seen in 
Figs.\ \ref{f.aanpscaling} and \ref{f.abscaling}. 

In Fig.\ \ref{f.icall} the BB case is actually well-known. In
Ref.\ \onlinecite{ullahfetter} it was calculated using a channel
geometry, rather than the empty half-spaces on the 
two sides. The fact that the values in this reference are slightly
larger may be just an indication of the restricted form of order parameter which was
assumed there. Thus we confirm the expectation that the critical
currents should only depend on the properties of the weak link
itself, not the way in which the current is driven through it. 

The other three panels of Fig.\ \ref{f.icall} show the interesting
feature that if there is A phase on one of the sides, the critical currents
will always approach the same values for large $D$. They are thus
determined by the inside of channel, and not the complicated
$\lvechat$ structures or the A-B interface at the ends of the junction.
The transition to the long-channel limit involves locking of $\lvechat$
perpendicular to the channel walls ($\lvechat=\pm\yvechat$) at the
middle of the junction, as in a parallel-plate geometry with plate
separation $W$.\cite{fetterullah}
For AA$\uparrow\downarrow$ this happens smoothly, but in
AA$\uparrow\uparrow$ and  AB there are rather sharp cusps
in the $J_c$ curve beyond which $J_c$ is
almost constant. In both cases this transition is due to a
crossing over of the critical currents of two branches of CPR with different
$\lvechat$ configurations. For $AA\uparrow\uparrow$ the configurations
are the symmetric (e) and the asymmetric (d or f) ones in
Fig.\ \ref{f.lconfs}, as discussed in the previous section. 
Some of the associated CPR's were shown in
Fig.\ \ref{f.aappd}, which corresponds to the $W/\xi_{GL}=6$ line of
Fig.\ \ref{f.icall} --- for small $\Delta\phi$ the symmetric $\lvechat$
configuration is the one to determine $J_c$. The $J_c$-transition lines
are shown in the phase diagram, Fig.\ \ref{f.pdiag}, although it should
be kept in mind that their positions may depend on the way in which
the numerics was done. 
For small $D$ the behaviors of all four cases are quite different. For
AA$\uparrow\downarrow$ and AB $J_c$ tends to be
suppressed, whereas for BB and AA$\uparrow\uparrow$ it 
increases. In AB there are actually $J_c$ transitions at two different
$D$ values. Both are due to switching of $J_c$ between the $0$ ($+$) or 
$\pi/2$ ($-$) CPR branches, which are seen in Fig.\ \ref{f.abscaling}. 
The $+/-$ regions are shown in Fig.\ \ref{f.pdiag}. Notice that 
all these details are only valid for $\dhat_\mu=\pm R_{\mu i}\shat_i$.
For other configurations the $F_J^{\rm tun}$ term is also finite, and 
the critical currents will in fact tend to be larger (for small $D$)
than presented in the figures above. 

Also shown in Fig.\ \ref{f.icall} are the critical currents for some
apertures of varying size, but fixed aspect ratio. These correspond
essentially to Figs.\ \ref{f.aanpscaling} and \ref{f.abscaling} for
$AA\uparrow\downarrow$ and AB, and to
equivalent (but not shown) figures for BB and AA$\uparrow\uparrow$.
Small numerical differences are due to the different computational
parameters. These curves are interesting
because all the temperature dependence in the GL regime is in
the length scale $\xi_{GL}(T)\sim(1-T/T_c)^{-1/2}$ alone, and scaling 
$D/\xi_{GL},W/\xi_{GL}$ with factor $l$ is thus
equivalent to changing the temperature: $l\sim(1-T/T_c)^{1/2}$. 
The aperture in the Paris experiments reported in Ref.\ \onlinecite{paris} was
a slit of dimensions $0.18\times2.6~\mu$m$^2$ in a $0.1~\mu$m thick
SiN membrane. The aspect ratio $W:D$ for this is close to $2:1$, which was
therefore most frequently used in our calculations. 
The behavior at small $l$ appears to follow a power law $J_c/J_{A,B}\sim l^\delta$.
For all of the three BB cases the exponent
is roughly the same, $\delta=1.8$. Similarly, for both AA$\uparrow\downarrow$
and AA$\uparrow\uparrow$ it is $\delta=2.2$, whereas for AB
$\delta=4.5$.
In BB the $J_c$ curves appear to be linear for large $l$, but this is
only illusory. 
To point out another numerical accident, note that in the
AA$\uparrow\uparrow$ case the critical current transition appears 
to occur at exactly the aspect ratio $W:D=2:1$. There seems to be no
obvious reason for this.

However, the striking similarity in the $J_c$ scaling behavior of 
the AA$\uparrow\downarrow$ and AA$\uparrow\uparrow$ configurations 
for small $l$ appears very surprising at a first glance. 
Since in the pinhole limit the critical current for AA$\uparrow\downarrow$
vanishes, should it not at least vanish faster as the smaller
dimensions are approached?
Two things should be realized here. The first is due to  the 2D nature of
our model: as the slit is made very
narrow, eventually all but the $\zvechat$ directional component
of the orbital part of the order parameter are suppressed.
The remaining coupling through a \emph{single} complex component
is direction-independent, and thus the behavior for both AA
cases becomes similar.
Second, even
for a 3D aperture, the ``pinhole'' behavior would not be exactly reached in
this limit, because the quasiclassical pinhole calculation relies on a
\emph{nonlocal} coupling through the ``$f$ functions''.
The present GL calculation is \emph{local}, and assumes that the
\emph{pairing potential} is everywhere nonzero inside the weak link ---
otherwise there could be no coupling at all. A different approach for 
a finite-size BB junction was recently adopted in Ref.\ \onlinecite{endres}.
These comments are also valid for the BB case, which explains the
somewhat different result of Section 5 as compared with Ref.\
\onlinecite{viljas2}


\section{CONCLUSIONS AND DISCUSSION} \label{s.discussion}

In the sections above we have presented a $p$-wave Ginzburg-Landau 
analysis of a two-dimensional weak-link model. We have studied the
current-phase relations and mapped the critical currents of the weak 
link, when it is placed between two bulk volumes of either superfluid
$^3$He-A or $^3$He-B. 
The BB case was studied before in several
restricted calculations,\cite{monien,ullahfetter,thuneberg88,viljas}
but here we were able to use much more general boundary conditions 
on the bulk phases. We showed that
similar ``$\pi$ states'' as in BB pinholes\cite{yippi,viljas2} exist
also for small finite-size apertures in the GL regime. They are due to the
second-order terms in Eq.\ (\ref{e.expansion}) and are thus associated
with small critical currents. On the other hand,
we showed how this behavior gradually becomes more complicated in
larger apertures, as the order parameter has more freedom to vary
inside the hole. In this way, various kinds of $\pi$ states may exist
together with either continuous or hysteretic CPR's which have
relatively large critical currents.
This is consistent with the experimental findings of Ref. \onlinecite{paris},
where, apparently due to the uncontrollability of the bulk textures, 
many kinds of CPR's were seen.
As in Ref.\ \onlinecite{viljas2} for the ``anisotextural'' effect in a
large pinhole array, it
was found that the hysteresis of CPR's tends to be smaller when the 
bulk boundary conditions are less symmetric. This enabled jumps to 
the $\pi$ states which were unreachable in symmetric cases.\cite{viljas}
Furthermore, we calculated the spin currents, and related
their behavior to transitions between different order-parameter
configurations in the aperture.
We also studied AA and AB junctions, which have previously not
been studied in any depth, at least not within the GL
theory for large apertures. We analyzed their symmetries, and related 
transitions in the
$\lvechat$ field to changes in forms of the CPR's and their critical
currents. In AB junctions, the usual phase-inversion symmetry was 
shown to be broken, which results in more complicated CPR's.
In addition to all these, a hydrostatic theory of the asymptotic 
phase corrections is presented in Appendix A, and an analysis of the 
numerical relaxation method is given in Appendix B.

We have also carried out some tests with the channel geometry
used in Ref.\ \onlinecite{ullahfetter}, and again we found the critical
currents to match with those in \mbox{Fig.\ \ref{f.icall}}. Also a geometry
with the weak link between two \emph{parallel} flow channels was tested.
It was found that the critical currents of a BB weak link are not
affected by the parallel flow before the flow itself goes unstable
by nucleating a vortex at some wall. 
However, these studies were not carried much further.
What we have also not done above is an analysis of the
effects of a strong magnetic field. The required magnetic GL 
terms would be relatively easy to include in the calculation, but they
would add more dimensions to the already large parameter space.
Actually there remain even some untested phase combinations which may
be possible for junctions with small $W/\xi_{GL}>3$ and $D\gg W$: a BAB or ABA
heterostructure. Such situations
may be obtainable at pressures close to
$p_0$. This is because the A phase tends to be slightly more stable in
restricted ``parallel plate'' geometries, and on the other hand
because there are some hysteresis effects related to the equilibrium
position of the A-B interface as the pressure is varied.\cite{viljasab}
The BAB configuration was recently discussed as an explanation of
some dissipation effects,\cite{dissipation} and this was, apparently,
the initial reason for considering the BAAB structure in Ref.\ \onlinecite{nishida}.
Although it should be possible to prepare such configurations in our program
in some pressure regimes, their stability in a real experiment
is questionable. Of course, the A phase could be stabilized more strongly
inside the aperture by using a localized magnetic field.
Also, since a self-consistent, general-temperature
calculation of a finite-size aperture (with strong-coupling effects included) is still
missing, no final conclusions can be drawn at this stage.
Some of these issues may be worth further studies at some later
time.



\section*{ACKNOWLEDGMENTS}
The Center for Scientific Computing (CSC) is acknowledged for
computer resources. A. Schakel, R. H\"anninen, and J. Kopu are
thanked for helpful discussions.

\appendix

\section{ASYMPTOTIC SOLUTIONS IN 2D} \label{s.asymptotics}

In this appendix we discuss the approximations used in the
asymptotic regions between $R_c$ and $R_\infty$ of Fig.\
\ref{f.halfregion}. The need for
these derives from a requirement to keep the numerical computation
at a minimum, but at the same time have the bulk boundary conditions
imposed as far away from the junction as possible. In 3D the boundary
conditions ($R_\infty$) could be taken all the way to infinity using
this method, but in 2D some smaller value has to be chosen. 
Since in this case $R_\infty$ is rather arbitrary, this procedure
is to some extent only cosmetic, enabling a smoother transition of the
order parameter at the $R_c$ cutoff. Thus, whenever this was not needed
(like in evaluating the critical currents), the asymptotic corrections
were not used.

At $R_\infty$ the boundary conditions are given by order parameters of the form
$A^{L,R}_{\mu i}=R_{\mu\nu}^{L,R}A^{(0)L,R}_{\nu i}(x)e^{\iu\phi^{L,R}}$
as shown above. However, since $R$ and $\phi$ are constant, these
states carry no mass or spin currents,
apart from the spontaneous spin currents parallel to B-phase surfaces.\cite{zhangwall} 
The order parameters must be corrected in the asymptotic regions in
order for them to conserve
the currents coming from the weak link aperture.
In this respect, the A and B phases can again be treated in
the same way, and the asymptotics on each side can be assumed to take the form:
\begin{equation} \label{e.asymptop}
A(\xvec)=\td R^s(\xvec) 
R A^{(0)}(x) [\td R^o(\xvec)]^T e^{\iu(\phi+\delta\phi(\xvec))}.
\end{equation}
Here $\delta\phi(\xvec)$ is a small phase correction and 
$\td R^s(\xvec)=R(\delta\theta_{\alpha}(\xvec))$ a small spin
rotation: $R^s_{\mu\nu}=\delta_{\mu\nu}-\epsilon_{\mu\nu\alpha}\delta\theta_\alpha$.
In addition we have included a small orbital rotation
$\td R^o(\xvec)=R(\delta\epsilon_{i}(\xvec))$. This is 
only needed in the anisotropic A phase to include corrections to the $\lvechat$ 
texture, which are present even in the absence of currents. 
In deriving the asymptotic corrections to $\delta\phi$,
$\delta\theta_\alpha$, $\delta\epsilon_i$ and the energy, we shall make the 
hydrodynamic approximation that $A^{(0)}$ is everywhere in its
bulk form. In the B phase this means that $A^{(0)}_{\mu i}=\Delta_B\delta_{\mu i}$
and for the A phase 
$A^{(0)}_{\mu i}=\Delta_{A}\delta_{\mu x}(\delta_{iy}\pm\iu \delta_{iz})$. 
For convenience we shall denote below the small-angle fields
$\delta\phi$, $\delta\theta_\alpha$, $\delta\epsilon_i$ with $\phi$,
$\theta_\alpha$, $\epsilon_i$, respectively. 

In this context we refer to all positions with respect to the
origin $(x,y)=(D/2,0)$. As a first approximation, all the small-angle
corrections will be assumed to depend only on the radial distance
$r=\sqrt{x^2+y^2}$ from this point, possibly with some simple
dependence on the azimuthal angle $\varphi=\arctan(y/x)$ also.


\subsection{Asymptotics in BW State} \label{ss.b}

We first deal with the case of B phase which is numerically simpler
to handle than the A phase, because it is isotropic (meaning that it
has an isotropic superfluid density tensor). There is no need
to worry about similar symmetry breaking effects as with the A phase
$\lvechat$ vector (see below), and \mbox{$\td R^o=1$}. On the other hand, the
``entanglement'' of the spin and
orbital parts of the order parameter makes their separation impossible, 
which complicates the analytic treatment. 
To facilitate the analysis of the gradient part of Eq.\ (\ref{e.fbulk}), we first change to a spin basis where the 
constant $R$ in Eq.\ (\ref{e.asymptop}) is replaced by a unit matrix. This is done by noting 
that $\td R^s R = R \td R^{s'}$, with
$\td R^{s'}_{\mu\nu}=\delta_{\mu\nu}-\epsilon_{\mu\nu\alpha}\theta_\alpha'$,
and $\theta_\alpha'=R^T_{\alpha\beta}\theta_\beta$, where the identity
$\epsilon_{ijk}R_{jl}R_{km}=R_{in}\epsilon_{nlm}$ is useful.
The spin current in this new basis is obtained with 
${j^{\spin}_{\alpha i}}'=R^T_{\alpha\beta}{j^{\spin}_{\beta i}}$.
However, to simplify notation we drop the primes below, remembering that
they should appear on all quantities with Greek spin indices.
With these definitions, the gradient 
energy (per unit length) can be written\cite{cross} 
\begin{equation} \label{e.bge}
F^B_G=\frac{1}{2}\int\upd^2x\big[ 
\rho_{\rmit{s}}\vvec_{\rmit{s}}^2 
+\rho^{\spin}_{\alpha\beta;ij}
{v^{\spin}_{\alpha i}}{v^{\spin}_{\beta j}} \big],
\end{equation}
and the mass and spin currents are
\begin{align}
\jvec_{\rmit{s}}&=\rho_{\rmit{s}}\vvec_{\rmit{s}} 
\label{e.bmc} \\
{j^{\spin}_{\alpha i}}&=\frac{\hbar}{2m_3}
\rho^{\spin}_{\alpha\beta;ij}{v^{\spin}_{\beta j}},
\label{e.bsc}
\end{align}
where $\rho_{\rmit{s}}=(2m_3/\hbar)^22(\gamma+2)K\Delta_B^2$,
$\rho^{\spin}_{\alpha\beta;ij}$
$=\rho_{\rmit{s}}[(\gamma+1)\delta_{ij}\delta_{\alpha\beta}$
$-(\gamma-2)\delta_{\beta i}\delta_{\alpha j}$
$-\delta_{\alpha i}\delta_{\beta j}]$,
$\vvec_{\rmit{s}}=(\hbar/2m_3)\boldsymbol\nabla\phi$ and
$\vvec^{\spin}_\alpha=(\hbar/2m_3)\boldsymbol\nabla\theta_\alpha$.
By making a variation in $\phi$ and $\theta_\alpha$ and noting the symmetry 
$\rho^{\spin}_{\beta\alpha;ji}=\rho^{\spin}_{\alpha\beta;ij}$  
we see that the extrema of $F^B_G$ are also found from the
continuity equations
$\boldsymbol\nabla\cdot\jvec_{\rmit{s}}$
$=\rho_{\rmit{s}}\boldsymbol\nabla\cdot\vvec_{\rmit{s}}$
$=(\hbar\rho_{\rmit{s}}/2m_3)\boldsymbol\nabla^2\phi=0$
and $\boldsymbol\nabla\cdot\jvec^{\spin}_{\alpha}=$
$\partial_i\rho^{\spin}_{\alpha\beta;ij}v^{\spin}_{\beta j}$
$=(\hbar\rho_{\rmit{s}}/2m_3)[(\gamma+1)\boldsymbol\nabla^2\theta_\alpha$
$-(\gamma-1)\partial_\alpha\partial_i\theta_i]=0$. The latter set of equations
can be directly diagonalized,\cite{vollhardt} but we assume the
solutions to be radial right from the beginning: $\phi(r,\varphi)=\td\phi(r)$ and 
$\theta_\alpha(r,\varphi)=\td\theta_\alpha(r)$.
Integration over angles in the energy [Eq.\ (\ref{e.bge})] and minimization 
with respect to these independent phase fields yields, for example, the 
solution
\begin{equation} \label{e.brad}
\phi(r,\varphi)=\td\phi(r)=\phi^c\frac{\ln(r/R_\infty)}{\ln(R_c/R_\infty)}.
\end{equation}
These satisfy $\td\phi(R_c)=\phi^c$ and $\td\phi(R_\infty)=0$.
Similar expressions are obtained for the spin phases $\theta_\alpha(r,\varphi)$.
In the $\rho_{\rmit{s}}$ units defined in Appendix B, the
energy corresponding to the solutions of the type of Eq.\ (\ref{e.brad}) is
\begin{align}
\td F^B_G&=\frac{2\pi}{4(\gamma+2)}\frac{1}{\ln(R_\infty/R_c)}
\Bigg\{ (\gamma+2)(\phi^c)^2 \nonumber\\ &+\sum_\alpha(\gamma+1) 
\left[1-\frac{1}{2}\frac{\gamma-1}{\gamma+1}(1-\delta_{\alpha z})\right]
({\theta_\alpha^c})^2 \Bigg\} 
\label{e.amptene}
\end{align}
and the asymptotic currents are
\begin{align}
\td J_{\rmit{s}}&=\pi \ln(R_\infty/R_c)^{-1}\phi^c
\label{e.bmcphi} \\
\td J^{\spin}_\alpha&=\pi(\gamma+1) \ln(R_\infty/R_c)^{-1} \nonumber \\
&\qquad\times
\left[1-\frac{1}{2}\frac{\gamma-1}{\gamma+1}(1-\delta_{\alpha z})\right] {\theta_\alpha^c},
\label{e.bsctheta} 
\end{align}
where no summation over $\alpha$ is implied.
By fitting the numerically calculated currents $\td J_{\rmit{s}}$, 
$\td J_\alpha^{\spin}$ (transformed with $R^T$to the
``primed'' spin basis) to Eqs.\ (\ref{e.bmcphi}) and (\ref{e.bsctheta}),
one can solve the parameters $\phi^c,\theta_\alpha^c$. Then one can
use Eq.\ (\ref{e.brad}) and the equivalent expression for $\theta_\alpha$ in
Eq.\ (\ref{e.asymptop}) to update the the asymptotic order parameter,
remembering first to transform $\theta_\alpha^c$ back into the
original ``unprimed'' spin coordinates.
There is an asymmetry in the spin parts of Eqs.\ 
(\ref{e.amptene}) and (\ref{e.bsctheta}):
the $z$ direction is slightly more rigid than the others. 
This results from the orbital-space (real-space) asymmetry of the
2D problem, which also affects the spin space in B phase. 
In the A phase this is not so, 
since there the spin and orbital parts of the order parameter
are decoupled; see below.

\subsection{Asymptotics in ABM State} \label{ss.a}

In the A phase 
we assume the following conditions for the vector 
$\lvechat=\mvechat\times\nvechat$:
(i) $\lvechat\cdot\wvechat=0$, where the orbital rotation axis 
$\wvechat$ is constant and in $yz$ plane, so that
(ii) $\wvechat\cdot\zvechat\approx 1$, 
and (iii) $\lvechat\cdot\xvechat=\cos\epsilon$ where the orbital rotation
angle $\epsilon$ is small.
>From condition (i) it follows that the phase $\phi$ is
well defined and acts as the superfluid velocity potential. Condition (ii) ensures that 
$\lvechat$ is essentially in the $xy$ plane with
$\vvec_{\rmit{s}}$, 
and (iii) simply means that the variations
from $\lvechat=\pm\xvechat$ are small.
Using a linearized approximation for $v^{\spin}_{\alpha i}$
the A phase gradient energy following from 
Eq.\ (\ref{e.fbulk}) can then be written 
\begin{equation} \label{e.age}
\begin{split}
F^A_G&=\frac{1}{2}\int\upd^2x\big[
\rho_{ij}v_{{\rm s}i}v_{{\rm s}j}
+\rho^{\spin}_{\alpha\beta;ij}
v^{\spin}_{\alpha i}v^{\spin}_{\beta j} \\
&+ (\hbar/2m_3)v_{{\rm s}i}C_{ij}(\boldsymbol\nabla\times\lvechat)_j 
+K_s(\boldsymbol\nabla\cdot\lvechat)^2 \\
&+K_b|\lvechat\times(\boldsymbol\nabla\times\lvechat)|^2
+K_t(\lvechat\cdot\boldsymbol\nabla\times\lvechat)^2 
\big]
\end{split}
\end{equation}
and the mass and spin currents are
\begin{align}
j_{{\rm s}i}&=\rho_{ij}v_{{\rm s}j} + (\hbar/2m_3)C_{ij}(\boldsymbol\nabla\times\lvechat)_j
\label{e.amc} \\
j^{\spin}_{\alpha i}&=\frac{\hbar}{2m_3}
\rho^{\spin}_{\alpha\beta;ij}v^{\spin}_{\beta j},
\label{e.asc}
\end{align}
where
$\rho_{ij}=\rho_\perp\delta_{ij}-(\rho_{\perp}-\rho_{\parallel})\lhat_i\lhat_j$, 
$\rho_\perp-\rho_\parallel=(\gamma-1)/(\gamma+1)$,
$\rho_{\perp}=(2m_3/\hbar)^22(\gamma+1)K\Delta_A^2$,
$\rho^{\spin}_{\alpha\beta;ij}=\delta_{\alpha\beta}\rho_{ij}$
$\vvec_{\rmit{s}}=(\hbar/2m_3)\boldsymbol\nabla\phi$, 
$\vvec^{\spin}_\alpha=$
$(\hbar/2m_3)(\dhat^\infty_\alpha\dhat^\infty_\beta$
$-\delta_{\alpha\beta})\boldsymbol\nabla\theta_\beta$,
$K_s=K_t=(\hbar/2m_3)^2\rho_\perp/(\gamma+1)$,
$K_b=(\hbar/2m_3)^2\rho_\perp\gamma/(\gamma+1)$ and
$C_{ij}=C_\perp\delta_{ij}-(C_{\perp}-C_{\parallel})\lhat_i\lhat_j$.
Here $\dhat^\infty_\alpha=R_{\alpha x}$ is the bulk spin vector.
The first three terms of Eq.\ (\ref{e.age}) are just 
$\vvec_{\rm s}\cdot\jvec_{{\rm s}}/2$ and 
$\vvec^{\spin}_\alpha\cdot\jvec^{\spin}_\alpha/2$, but
there are three additional terms resulting from the $\lvechat$ texture
alone. Here the actual value of the $C$ tensor will not be important to us, 
since with our assumptions the current component resulting from
$\boldsymbol\nabla\times\lvechat$ is either divergenceless ($\lvechat$ in plane)
or even vanishes ($\lvechat$ constant). As a result, we shall neglect
the $C$ terms also from the free energy, but keep all others initially. 

We now seek to  minimize the energy [Eq.\ (\ref{e.age})] with respect to the
small-angle fields $\phi$, $\theta_\alpha$ and $\epsilon$. In
principle these should be optimized simultaneously, but we shall
separate the problems so that in optimizing $\phi$ or $\theta_\alpha$ we
assume $\lvechat=\pm\xvechat$, and in correcting $\lvechat$ with
$\epsilon$ we assume that the velocities vanish. 
Assume, then, that we have $\lvechat=\pm\xvechat$, in which case the last
three terms in Eq.\ (\ref{e.age}) vanish.
By making a variation in $\phi$ and $\theta_\alpha$ and noting the symmetry 
$\rho^{\spin}_{\beta\alpha;ji}=\rho^{\spin}_{\alpha\beta;ij}$  
we see that the extrema of the first two terms of $F^A_G$ are found simply from the
continuity equations
$\boldsymbol\nabla\cdot\jvec_{\rmit{s}}=\partial_i\rho_{ij}v_{\rmit{s}j} = 0$
and $\boldsymbol\nabla\cdot\jvec^{\spin}_{\alpha}=$
$\partial_i\rho^{\spin}_{\alpha\beta;ij}v^{\spin}_{\beta j}=0$.
Now if we change to new coordinates 
$x'=x\sqrt{(\gamma+1)/2}$, $y'=y$, then the superfluid density
tensor $\rho'_{ij}$ will appear isotropic. In these coordinates the
continuity equations are simply the Laplace equations
$\boldsymbol{\nabla'}^2\phi=\boldsymbol{\nabla'}^2\theta_\alpha=0$. We attempt to solve these 
in polar coordinates $r',\varphi'$ of the primed system, where the
general solution is of the form
\begin{align}
\phi(r',\varphi')&=A\varphi'+B\ln(r'/a) \nonumber\\ +& \sum_{n=1}^{\infty}
(C_n {r'}^n+D_n {r'}^{-n})\sin(n\varphi'-\alpha_n).
\end{align}

The solution we are looking for is determined by setting boundary conditions on the $R_c$
cutoff and at infinity. We assume that the currents coming from the junction flow
radially as $r'\rightarrow\infty$. This is described by
the solution proportional to $\ln r'$. However, because the $\lvechat$
texture tends to align perpendicular to surfaces, it usually
becomes nonsymmetrical close to the weak link (see Section 5). 
Since the superfluid density is larger perpendicular to $\lvechat$, most of the current
will flow perpendicular to it. Therefore, the currents do not
flow in and out of the junction mirror-symmetrically with respect to the $xz$
plane.
\begin{figure}[!tb]
\begin{center}
\includegraphics[width=0.65\linewidth]{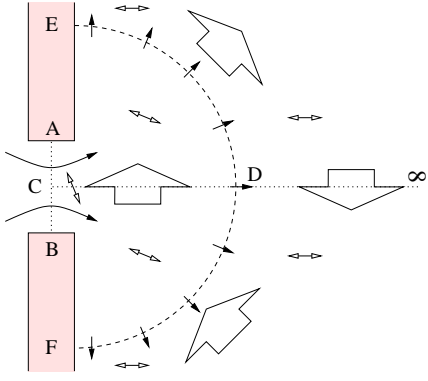}
\caption{Schematic representation of a circulating current 
component (large arrows) in A phase due to the bending of the orbital 
anisotropy axis $\lvechat$ (two-headed arrows). The small black arrows
denote the symmetric, radially decaying current component.} \label{f.ccirc}
\end{center}
\end{figure}
Such a flow field can be thought to consist of two components: one
which is symmetrical and radial, and another which circulates in front
of the opening as illustrated in Fig.\ \ref{f.ccirc}. For the
circulating part of the mass current we test two boundary conditions: 
$\partial_{r'}\phi(R_c)\propto\sin\varphi'$ and $\partial_{r'}\phi(R_c)\propto\sin2\varphi'$.
Both of these describe a flow which is outward above the $x$
axis when it is inward below it and vice versa, but the distributions
differ. The corresponding solutions are $\phi=\sin\varphi'/r'$ and
$\phi=\sin 2\varphi'/{r'}^2$ and the total velocity potential is thus of the
form
\begin{equation} \label{e.apot}
\phi(r',\varphi')=\phi^c\frac{\ln (r'/R_\infty)}{\ln(R_c/R_\infty)} 
+ \phi^{c,a} \frac{R_c^n}{{r'}^n}\sin(n\varphi'), ~ n=1,2.
\end{equation}
Here the coefficient $\phi^c$ is again determined by equating the current
obtained from Eq.\ (\ref{e.amc}) by integration over $R_c$ (surface
EDF in Fig.\ \ref{f.ccirc}) with the total current $J_{\rmit{s}}$ calculated
numerically at the junction (surface AB). The coefficient $\phi^{c,a}$, on the
other hand, is trickier. It could be determined by equating the current resulting from the
second term of Eq.\ (\ref{e.apot}) over the upper part of the cutoff
(surface DE)  with the
$y$-directional current $J^a_{\rmit{s}}$ calculated numerically over
the surface CD.
Note that in the case $n=1$ the circulating parts of the  
currents over surfaces DE, D$\infty$ and DF are equal,
and hence no net $y$ directional current is introduced. This is
not so for $n=2$, and, in fact, in this case the phase field gives an
unphysical-looking phase gradient perpendicular to the wall.
However, our approximation does not take into account
the fact that the order parameter is suppressed at the wall and 
(at least for a diffusive surface) the superfluid density vanishes
there. Therefore no \emph{current} should flow through it. 
The phase correction of Eq.\ (\ref{e.apot}) should only be seen as a
variational ansatz, and in practice we find that $n=2$ gives a 
better overall fit with lower total energies than $n=1$.
Similar expressions and conclusions exist for the spin velocity
potentials $\theta_\alpha$.

We now give the results for the current vs. phase correction
relationships. Let us remind that in deriving these we have assumed
that $\lvechat=\pm\xvechat$ in the asymptotic region. 
In the $\rho_\perp$ units defined in Appendix B, 
the gradient energy of the asymptotic solution is
\begin{align}
\td F^A_G=&\frac{1}{2}\pi\sqrt{\frac{2}{\gamma+1}}
\Bigg\{
\frac{(\phi^c)^2}{\ln(R_\infty/R_c)} 
+\sum_\alpha\frac{(\theta_\alpha^c)^2}{\ln(R_\infty/R_c)} \nonumber\\
&+\frac{1}{2n}[1-(R_c/R_\infty)^{2n}](\phi^{c,a})^2 \nonumber\\
&+\sum_\alpha\frac{1}{2n}[1-(R_c/R_\infty)^{2n}](\theta_\alpha^{c,a})^2
\Bigg\}
\end{align}
while the currents are
\begin{align}
\td J_{\rmit{s}}&=\pi
\sqrt{2/(\gamma+1)} \ln(R_\infty/R_c)^{-1}\phi^c ð
\label{e.amcphi} \\
\td J^{\spin}_\alpha&=\pi \sqrt{2/(\gamma+1)} \ln(R_\infty/R_c)^{-1}\theta_\alpha^c
\label{e.asctheta} \\
\td J^a_{\rmit{s}}&=
n\sqrt{2/(\gamma+1)}\phi^{c,a} ð
\label{e.aasymcphi} \\
\td J^{{\spin},a}_\alpha&=n\sqrt{2/(\gamma+1)}\theta_\alpha^{c,a}.
\label{e.aasysctheta}
\end{align}
Again, by fitting the numerically calculated currents to these expressions,
one can solve the parameters $\phi^c$, $\theta_\alpha^c$.
The above methods of estimating the ``asymmetric'' currents are very
unreliable, and in practice the corrections $\phi^{c,a}$ and
$\theta_\alpha^{c,a}$ were not used. They are presented here for
completeness.

In addition there is the small correction to the direction of
$\lvechat$, which is mostly due to the complicated surface
structure. There is also a small tendency for the $\lvechat$ to align
with $\pm\vvec_{\rmit{s}}$ at high superflow velocities, but
here we assume $\vvec_{\rmit{s}}=\vvec^{\spin}_\alpha=0$. 
We do not require $\lvechat$ to vary only in $xy$ plane, as
long as it is rotates only around the fixed axis $\wvechat$ which is in the $yz$ plane. 
The minimization of the last two terms of Eq.\ (\ref{e.age}) is done by assuming that 
$\lvechat(r',\varphi')=\pm R(\wvechat,\epsilon(r',\varphi'))\xvechat$, 
so that $\lvechat=\pm\xvechat$ for $r'\rightarrow\infty$ but also at the walls 
($\varphi'=\pm\pi/2$). This can be described with the
variational ansatz
$\epsilon(r',\vartheta')=\td\epsilon(r')\cos\varphi'$. 
If we further assume that $(K_{s,t}+K_b)/2\ll K_b(\gamma+1)/2$ in order
to drop some terms with $y$ and $z$ derivatives, the energy density 
resulting from the bending of $\lvechat$ is just 
$2f_g^l=K_b[(\partial_x\epsilon_y)^2+(\partial_x\epsilon_z)^2]$,
where $\epsilonvec=\epsilon\wvechat$. 
Minimization of this with the above ansatz gives
$\td\epsilon(r')\approx\td\epsilon^c(R_c/r')$, and the energy
\begin{equation}
\Delta \td F^A_G=\frac{3\pi\gamma}{16}\sqrt{\frac{2}{\gamma+1}}
[1-(R_c/R_\infty)^2] (\td\epsilon^c)^2. 
\end{equation}
Here $(\td\epsilon^c)^2=(\td\epsilon_y^c)^2+(\td\epsilon_z^c)^2$, and
$\td\epsilon_{y,z}^c\approx\lhat_{y,z}^c(x'=R_c,y'=0)$, 
These cannot be obtained from any
conservation law, but instead we extract them from the numerical
solution by requiring
continuity of $\lhat_y$ and $\lhat_z$ and their $x$ derivatives along the $x$ axis.
The $\lvechat$ vector can be obtained numerically from the $A$ matrix
inside the $R_c$ region with 
$\lhat_i=\epsilon_{ijk}\im(\td A_{\mu j}^*\td A_{\mu k})/2$, as shown in
Fig.\ \ref{f.lfields}. 


\section{NUMERICAL METHODS} \label{s.numerical}

This appendix considers in some detail the implementation of our
numerical relaxation methods. We skip the easier part of calculating
the 1D order parameter $A^{(0)}(x)$ for a planar wall, since
it is only needed as a boundary condition for the 2D calculation.
Of course, the same techniques can be used for that also, but
in practice the number of variables in the 1D minimization is so small
that perfecting the method is not of vital importance.
At all solid surfaces, we only consider the diffuse-scattering boundary
condition $A_{\mu i}=0$, see Section 3. 
Below we assume that the boundary conditions on the minimization
region are fixed, and comment on the addition of the asymptotic
corrections (Appendix A) in the last subsection. For the purpose of numerics, we
must consider all quantities to be dimensionless, and we first discuss
the unit reductions.
\subsection{Units}

A natural way to scale any quantity $L$ with the units of length is
$\td L = L / \xi_{GL}$, and a dimensionless order parameter is obtained
with $\td A = A /ð\Delta_{A,B}$, where $\Delta_{A,B}$ are 
the bulk gaps of A and B phases, respectively.
In the A phase $\Delta_A^2=\alpha/[2(2\beta_{245})]$, where
$\beta_{245}=\beta_2+\beta_4+\beta_5$, and 
for the B phase
$\Delta_B^2=\alpha/[2(3\beta_{12}+\beta_{345})]$, where
$\beta_{12}=\beta_1+\beta_2$, 
$\beta_{345}=\beta_3+\beta_4+\beta_5$.
Physically motivated units for the energies and currents are based on the 
the bulk superfluid densities
$\rho_{\perp}=(2m_3/\hbar)^22(\gamma+1)K\Delta_A^2$
and 
$\rho_{\rmit{s}}=(2m_3/\hbar)^22(\gamma+2)K\Delta_B^2$ 
(see appendix A).
Using these we define the following units for the currents and energy
(per unit length). In A phase
$J_A=(\hbar/2m_3)\rho_\perp$, 
$J^{\spin}_A=(\hbar/2m_3)^2\rho_\perp$, and
$E_A=(\hbar/2m_3)^2\rho_\perp$.
For B phase
$J_B=(\hbar/2m_3)\rho_{\rmit{s}}$, 
$J^{\spin}_B=(\hbar/2m_3)^2\rho_{\rmit{s}}$, and
$E_B=(\hbar/2m_3)^2\rho_{\rmit{s}}$.
It is these units which are used in all the figures of this paper, 
and sometimes we denote 
$\td J_{\rmit{s}}=J_{\rmit{s}}/J_{A,B}$, 
$\td J^{\spin}_\alpha=J^{\spin}/J^{\spin}_{A,B}$, 
and $\td F=F/E_{A,B}$.
Note that for weak coupling (see Section 3), the A and B phase superfluid densities 
$\rho_s$ and $\rho_\perp$ are equal, and so are the units.
For cases where both A and B phase are involved simultaneously, 
we choose to use the B phase units.

However, for numerics, a more convenient reduction of the
energy is given by
$\ovl f=f/[\alpha\Delta_{A,B}^2/2]$, and then
the reduced GL parameters are given by 
$\ovl\beta_ i=(2\Delta_{A,B}^2/\alpha)\beta_i$,
$\ovl\alpha=2$ and $\ovl K=2$.
Using the $\Delta_{A,B}$ we thus have in the A phase
$\ovl \beta_i=\beta_i/(2\beta_{245})$ and in the B phase
$\ovl \beta_i=\beta_i/(3\beta_{12}+\beta_{245})$.
In the formulas below, these unit reductions are assumed to have been 
performed, but we do not write any tildes or overlines on the symbols.

\subsection{The Minimization Problem}

The 2D numerical scheme is based on an optimization of the free energy
$F_{\Omega}$ [Eq.\ (\ref{e.geneene})] or, which is more-or-less
equivalent, solution of the corresponding 
nonlinear Euler-Lagrange field equations. 
In fact, the method to be
discussed can be applied (at least approximately) for solving a more general class of nonlinear
equations which have the form $G_{\mu i}(A(\xvec))=0$, where
$\mu,i=x,y,z$.
Nevertheless, we shall consider the the problem as that of
optimizing (minimizing) the GL energy functional $F_\Omega[A]$ in the region
$\Omega$ of Fig.\ \ref{f.halfregion}. More mathematically speaking, for the
given fixed boundary conditions $A^{L,R}$ 
[of the form of Eq. (\ref{e.asymptop})] on $\Gamma^{L,R}$, we wish to find 
$F_{J}[A^{L},A^R]$
$=\min_{A}F_\Omega[A]$.
As discussed in Appendix A, the full numerical
minimization is done only inside the region bounded by the $R_c$
cutoffs, but the computational region can be effectively increased
by making use of the asymptotic corrections between $R_c$ and
$R_\infty$. 


In the following, all entities $A$ with the $3\times3$ matrix structure
$A_{\mu i}$ where $\mu,i=x,y,z$ will be referred to as \emph{spin-orbit
matrices}, or \emph{spin-orbit fields}. For simplicity of notation,
let us define the following rotational invariants which are functions
of the spin-orbit matrices $A,B,C,D$:
\begin{align} 
f_{\alpha}(A,B) & = \re \Tr(AB\TC), \label{e.invalpha} \\
f_1(A,B,C,D) & = \re \Tr(A\C\!B\TC) \Tr(CD\T), \label{e.inv1} \\
f_2(A,B,C,D) & = \re \Tr(AB\TC)\Tr(CD\TC), \label{e.inv2} \\
f_3(A,B,C,D) & = \re \Tr(AB\T\!C\C\!D\TC), \label{e.inv3} \\
f_4(A,B,C,D) & = \re \Tr(AB\TC\!CD\TC), \label{e.inv4} \\
f_5(A,B,C,D) & = \re \Tr(AB\TC\!C\C\!D\T), \label{e.inv5} 
\end{align}
\begin{align}
g_1(A,B) & = \re \partial_{i}A_{\mu i}^*\partial_{j}B_{\mu j}, 
\nonumber \\ 
g_2(A,B) & = \re \partial_{i}A_{\mu j}^*\partial_{i}B_{\mu j},
\nonumber \\ 
g_3(A,B) & = \re \partial_{i}A_{\mu j}^*\partial_{j}B_{\mu i}. \label{e.invK3}
\end{align}
The invariants have several symmetries, some of which are listed below:
\begin{align} 
f_{\alpha}(A,B)&=f_{\alpha}(B,A)=f_{\alpha}(A^*,B^*), \\
f_{1,2}(A,B,C,D)&=f_{1,2}(A,B,D,C), \\ 
f_{3,4,5}(A,B,C,D)&=f_{3,4,5}(B,A,D,C), \\ 
f_{1-5}(A,B,C,D)&=f_{1-5}(C,D,A,B), \\
f_{1-5}(A^*,B^*,C,D)&=f_{1-5}(A,B,C^*,D^*), \\
g_{1,2,3}(A,B)=g_{1,2,3}&(B,A)=g_{1,2,3}(A^*,B^*). \label{e.gsymm}
\end{align}
It is worth noting here that symmetry of Eq.\ (\ref{e.gsymm}) of the
gradient terms is not necessarily valid in a discretized form, unless symmetric
difference approximations are used for all derivatives.
Using the definitions of Eqs.\ (\ref{e.invalpha}-\ref{e.invK3}), the GL energy
functional [Eqs. (\ref{e.geneene}) and (\ref{e.fbulk})] can now be rewritten as follows
\begin{align} \label{e.invene}
F[A,A^*]=&\int\upd^3x\big[ 
-\alpha f_{\alpha}(A,A) + \sum_{i=1}^3K_ig_i(A,A) \nonumber\\
&+\sum_{i=1}^5 \beta_if_i(A,A,A,A)  \big].
\end{align}
For the purpose of numerics, it is very helpful to 
consider $F$ as a functional of both $A$ and $A^*$ independently, as
shown here explicitly.
Note that the third gradient term
in (\ref{e.invK3}) can be integrated by parts and added to the first
term, if a separate surface energy term is introduced. 
Then, if the surface and the associated boundary conditions are
symmetric enough, the surface term may vanish altogether, thereby
reducing the numerical effort somewhat. Since this is not always the
case, we shall not assume it anywhere.

\subsection{Line Searching in Minimization}

The basic component of any efficient minimization algorithm is a
method for doing \emph{line searches}, \emph{i.e.}, doing one-dimensional 
minimizations of the function in some given direction. More precisely,
one wishes to minimize
\begin{equation}
h(\lambda)=F[A+\lambda D,A^*+\lambda D^*]
\end{equation}
with respect to the real parameter $\lambda$, given the ``starting
point'' $A$ and a ``search direction'' $D$. The extrema of
$h(\lambda)$ are found from the zeros of its derivatives, \emph{i.e.}, by solving
\begin{equation} \label{e.hprime}
h'(\lambda)=2 \re \left( G(A+\lambda D),D\right)=0,
\end{equation}
where $G(A(\xvec))=\delta F[A,A^*]/\delta A^*(\xvec)$, and
we have defined a \emph{scalar product} of two spin-orbit fields $A$ and $B$ as
follows:
\begin{equation} \label{e.scalar}
(A,B)=\int\upd^3x\Tr(A(\xvec)B\!\TC(\xvec)).
\end{equation}
It is easy to check that this is indeed bilinear, satisfies
$(B,A)=(A,B)^*$, and that $(A,A)\geq 0$, with $(A,A)=0$ only for
$A=0$. The scalar product
also gives a natural definition of a \emph{norm} for the
spin-orbit field: $||A||=\sqrt{(A,A)}$. 
Note that, with respect to the
matrix indices, this is the so-called \emph{Frobenius norm}.

\begin{figure}[!tb]
\begin{center}
\includegraphics[width=0.75\linewidth]{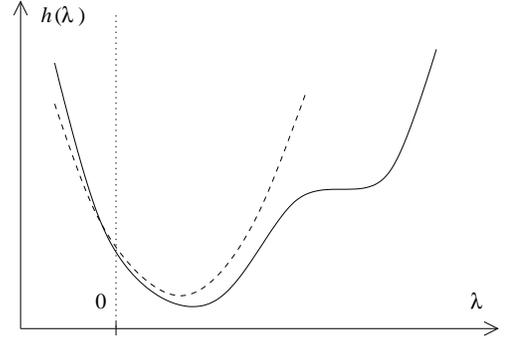}
\caption{A fourth order polynomial (solid line) and a local quadratic
approximation to it (dashed line) around $\lambda=0$ (dotted line).} \label{f.poly}
\end{center}
\end{figure}

The interesting point is that, since the GL free energy is only of
fourth order (quartic) in the order parameter, the function $h(\lambda)$ is
simply a fourth order polynomial with real coefficients
\begin{equation}
h(\lambda)=a\lambda^4+b\lambda^3+c\lambda^2+d\lambda+e.
\end{equation}
The fourth-order coefficient is clearly always positive, so that the graph of
the function must look like the curve in Fig. \ref{f.poly}.
The number of minima is 1 or 2, but only those with
$\lambda>0$ will be of interest here.
If one is now willing to take the small trouble of calculating the
coefficients $a$ to $e$, then the \emph{exact} minima of $h(\lambda)$ are
immediately found by solving the real roots of a third-order
equation $h'(\lambda)=0$. 
For the conjugate gradient method the accuracy of line searches
is important, at least in theory, and we list below the
formulas for $a$ to $e$ for the GL free energy discussed above:
\begin{equation} \label{e.a}
a=\int\upd^3x[\sum_{i=1}^5\beta_if_i(D,D,D,D)] 
\end{equation}
\begin{equation}
b=\int\upd^3x[4\sum_{i=1}^5\beta_if_i(D,D,D,A)] 
\end{equation}
\begin{align}
c&=\int\upd^3x\{2\sum_{i=1}^5\beta_i[f_i(D,D,A,A)+f_i(D,A,D,A)
\nonumber \\
& \quad+f_i(D,A,A,D)]-\alpha f_{\alpha}(D,D) + \sum_{i=1}^3 K_ig_i(D,D) \}
\end{align}
\begin{align} \label{e.d}
d=&\int\upd^3x\{4\sum_{i=1}^5\beta_i f_i(D,A,A,A) 
-2\alpha f_{\alpha}(D,A) \nonumber \\ 
&+ \sum_{i=1}^3 K_i[g_i(D,A)+g_i(A,D)] \}
\end{align}
\begin{equation}
e=F[A,A^*]
\end{equation}
Here the last term of $d$ can now be simplified with 
Eq.\ (\ref{e.gsymm}), but only with the proviso 
that the difference approximations are symmetric.
The coefficient $e$ (free energy at position $A$) is just a constant
and unnecessary for the line searching procedure. 

The numerical calculation can be further simplified by assuming 
the coefficients $a$ and $b$ to be zero. This is equivalent to
a local quadratic approximation of the energy functional, which 
is reasonable if the center point $A$ is already
close to the minimum of $F$ (see Fig.\ \ref{f.poly}).
Note that Eqs.\ (\ref{e.a}-\ref{e.d}) for the coefficients
$a$ to $d$ can also be obtained directly from
Eq.\ (\ref{e.hprime}), given the GL equations 
$G(A)=0$. This is indeed the only
way, if no such functional exists whose Euler-Lagrange equation
this $G(A)=0$ would be. Of course, in that case there is no
guarantee that the CG iteration will converge at all.

\subsection{The Conjugate Gradient Method}

Perhaps the simplest approach for the minimization of a nonlinear 
function(al) is the \emph{steepest descent} (SD) method. In this method, on each ($k$th)
step of iteration, the line search direction is simply chosen to be
that of the negative gradient: $D_k=-G_k$, where $G_k=G(A_k)$. The line search condition
[Eq.\ (\ref{e.hprime})] then implies that $(D_k,D_{k-1})=0$, \emph{i.e.}, successive
search directions are perpendicular to each other with
respect to the appropriate scalar product [Eq.\ (\ref{e.scalar})].
This leads to the well-known zigzag-type trajectory, which in
many cases  results in extremely slow convergence.\cite{bazaraa} The situation
can usually be helped somewhat by skipping the costly line search procedure
altogether, by choosing $A_{k+1}=A_{k}-c\cdot G_k$, where $c$ is
a given small positive constant. This procedure is in fact equivalent to
a simple (``forward Euler'') time-discretization of the time-dependent GL (or TDGL)
equations, where $c$ is proportional to the time step and the strength
of dissipation.\cite{thuneberg87} By following the progress of this iteration one may thus
get a rough idea of the dynamics of the order parameter when, for example,
a phase slip in a weak link occurs.

At least in the case of the GL equations, a significant
speedup of convergence can be achieved by taking advantage of the
knowledge of previous search directions. In the standard
\emph{conjugate gradient} (CG) methods\cite{bazaraa} the iteration step $k$ is taken as follows:
\begin{quote}
\textbf{Conjugate gradient iteration step} 
\begin{itemize}
\item $D_k=-G_k + \beta_k\cdot D_{k-1}$ 
\item Do a line search of 
$h(\lambda)=F[A_k+\lambda D_k,A_k^*+\lambda D_k^*]$ 
to find its (smallest) positive minimum point  $\td\lambda$
\item Set $A_{k+1}=A_k+\td\lambda\cdot D_k$
\end{itemize}
\end{quote}
The constant $\beta_k$ can be calculated in one of several ways.
Two of the most commonly used are the \emph{Fletcher-Reeves} choice
$\beta_k=(G_k,G_k)/(G_{k-1},G_{k-1})$,
and the \emph{Polak-Ribi\`ere} choice
$\beta_k=(G_k-G_{k-1},G_k)/(G_{k-1},G_{k-1})$.
The latter can be slightly more efficient, but it has the
memory-economical disadvantage that the gradient $G_{k-1}$ of 
the previous round must also be stored. The method is usually started
with $D_0=-G_0$ which is guaranteed to be a direction of
descent. During the iteration the process can also be restarted every
once in a while. This is because it is possible that inefficient
search directions will begin to be generated
after a few rounds of minimization, if the function(al) deviates very strongly from a 
quadratic form. Restarting should be done at least if the line search fails
to find a positive $\td\lambda$, and possibly on every $N$th step or so,
where $N$ is the total number of real variables in the discretized 
order-parameter field.\cite{bazaraa}

It seems that in practice neither the accuracy of the line search, nor
the choice for the $\beta_k$'s is very important, at least if the
method is sometimes restarted. The most important
part is really the basic philosophy of using the previous search
direction as described above.

One of the benefits of exact line searching is that it provides the small
\emph{change} in energy between steps very naturally through $\Delta F_k$
$=a\td\lambda^4+b\td\lambda^3+c\td\lambda^2+d\td\lambda$, so that 
$e$, the energy itself, does not need to be calculated on every round. 
This appears to be a more accurate way to obtain the change than application of 
$\Delta F_k = F[A_k,A_k^*]-F[A_{k-1},A_{k-1}^*]$ after completion of
each iteration step. 
This is an important point when the limit of available floating point accuracy is
eventually approached: one should always have $\Delta F_k < 0$
for a \emph{descent} direction $D_k$, and it is wise to check that this
is indeed the case.
However, a mistake would be done by assuming that 
$F[A_n,A_n^*]=\sum_{k=0}^{n-1}\Delta F_{k}+F[A_0,A_0^*]$, since this
will involve summation of numbers whose orders of magnitude are
very different from each other. Sufficiently close to the minimum an $F$ calculated in
this way may no longer change at all during the iteration, even if the the gradient is
still nonzero. The change in energy should therefore be used as a judge of
convergence only with care.

\subsection{Comparison of Methods}

\begin{figure}[!tb]
\begin{center}
\includegraphics[width=0.9\linewidth]{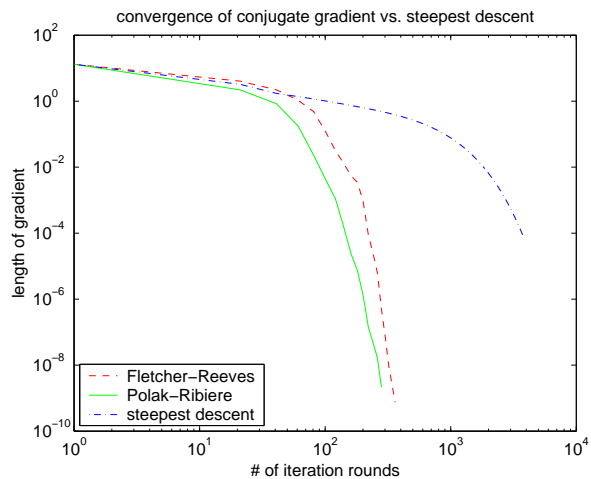}
\caption{Gradient length versus number of finished iteration rounds on
a log-log scale. The solid and dashed lines are for the
Polak-Ribi\`ere and Fletcher-Reeves variants of the
conjugate gradient method, respectively, and the
dash-dotted line for the steepest descent method. All methods use exact
line searches, so that the real time spent on one round is practically the same
for all of them. For a tolerance level of $10^{-4}$ --- where the
steepest descent method was terminated by hand --- greater than
order-of-magnitude differences in numbers of required iteration steps
are visible. The conjugate gradient methods were restarted on round
200, which did not affect the convergence in any noticeable way.} \label{f.comparison}
\end{center}
\end{figure}

Here we compare briefly the convergence of the different methods.
We use an example calculation in the geometry of Fig.\ \ref{f.halfregion},
with $R_c/\xi_{GL}=25$, $W/\xi_{GL}=10$, 
$D/\xi_{GL}=6$, with no
asymptotic corrections ($R_\infty=R_c$), and B phase 
with $\omegavechat^L=\omegavechat^R$ on both sides. 
The discretization lattice spacing was chosen to be 
$\Delta x/\xi_{GL} = \Delta y/\xi_{GL} = 1$.
Figure \ref{f.comparison} shows plots of the length of the gradient on
the $k$th iteration round, $||G_k||=||G(A_k)||$, versus the round
number $k$. Both the Fletcher-Reeves and Polak-Ribi\`ere variants
of the CG method are compared to the SD method. The superiority of
either form of CG
is evident from this figure: while the CG methods reach the level
$||G_k||\approx10^{-4}$ for $k=200\ldots300$, the SD method requires 
almost $k=4000$ for it. The real time needed by SD is thus more than
tenfold compared to CG. 
This does not reveal anything
rigorous about the	convergence classes, but for the 
CG methods the convergence must be superlinear, and probably close 
to quadratic.\cite{bazaraa} The scaling properties (behavior of convergence
with increasing problem size) were also not tested. 
As an interesting detail, it was noticed that the A phase tended
to converge faster than B phase.

Another possibility for obtaining faster convergence compared to the
SD algorithm would be to use \emph{quasi-Newton} (secant) methods,
which utilize knowledge of the Hessian matrix or some approximation to
it.\cite{bazaraa} However, in our
example problem the CG method has several advantages over
these methods also: (i) While analytic calculation of the second
derivatives of $F$ is possible, the Hessian matrix can easily
become huge and take up a lot of memory. This is not so severe in
approximate schemes (such as BFGS) which do not store the whole
Hessian matrix. (ii) The Newton method
has an inconvenient property of being unstable, if the initial guess
is not chosen close enough to the minimum. To avoid problems, special
stability checks are needed. (iii) Most inconveniently of
all, the construction of the Hessian matrix and solving the related linear system of
equations would require the programmer to \emph{order} the field variables in
some arbitrary way to form a \emph{vector}. In the case of a computational region looking like
the one in Fig.\ \ref{f.halfregion} and not a rectangle, there is no obvious way to do such
ordering. Any choice would require one to program obscure and lengthy
routines for transformations between the ``$xy$'' and vector orderings.
The CG method requires no such thing, and all field
variables can be iterated in any order on each iteration step,
independently of each other.
(iv) The explicit use of $A$ and $A^*$ as the independent variables
instead of $\re A$ and $\im A$ is also simplest in CG type methods. The
practical necessity of this is clear from the complicated structure of the free
energy functional. Points (iii) and (iv) are also why an application of most ready-made
library routines is not very attractive for problems in general $p$-wave
Ginzburg-Landau theory. 


\subsection{Updating the Asymptotic Values}

Finally we consider the task of fitting
the asymptotic solutions of Appendix A to the numerical solutions 
obtained by the above methods for $A^{L,R}$ fixed 
on $r=R_c$.
This could be done in several ways, but here we choose a very
simple one, and describe it only in therms of the B phase. 
The basic idea is iterative. Start with $\phi^c=\theta_\alpha^c=0$ in 
Eq.\ (\ref{e.brad}), an thus no phase corrections in the
the asymptotic order parameter [Eq.\ (\ref{e.asymptop})].
Then (i) minimize $F$ inside the $R_c$ cutoffs, (ii) calculate the
currents $\td J_{\rmit{s}}$
and $\td J^{\spin}_\alpha$ in the junction, (iii) insert them in
Eqs.\ 
(\ref{e.bmcphi}) and (\ref{e.bsctheta}) to get new $\phi^c$ and
$\theta_\alpha^c$, (iv) update Eq.\ (\ref{e.asymptop}) accordingly, and then
repeat from (i) until satisfactory conservation of the
currents over the $R_c$ cutoff is achieved. 
If we denote by ``out'' the current in the asymptotic region,  
and $\Delta \td J_{\rmit{s}}\equiv$
$\td J_{\rmit{s}}^{\rmit{out}}-\td J_{\rmit{s}}$
then the requirement was usually
$|\Delta \td J_{\rmit{s}}|\leq10^{-4}$, and similarly for the
spin currents.

Note that 
$\Delta \td J_{\rmit{s}}\propto$
$\partial \td F_{\rmit{\Omega}}/\partial\phi^c$, where $\td
F_{\Omega}$ is the energy in the whole $\Omega$ for given 
$\phi^c, \theta_\alpha^c$, and thus the process (i-iv) is seeking the
minimum of $\td F_{\Omega}$.
Since this is obviously a
very inefficient method, the benefit of fast convergence by
the conjugate gradient method in the $r<R_c$ part of $\Omega$ is lost if this
iteration is continued very far. Fortunately, the reason for
taking the corrections into account is mostly cosmetic in practice, and
only a few (less than 10) iteration rounds may already lead to a reasonably
continuous order parameter and current across the $R_c$ cutoff.
In the case of A phase, also the correction of
$\lvechat$, and possibly the asymmetric phase corrections,
should be updated on each round. See Appendix A for details.

\end{document}